\documentstyle[12pt]{article}
\newcommand{\nsigma}{\mbox{\boldmath $\sigma$}}
\newcommand{\ntau}{\mbox{\boldmath $\tau$}}
\newcommand{\onehalf}{\frac{1}{2}}
\begin{document}
\title{   {\bf Model calculations of doubly closed shell nuclei in
CBF theory} \\ 
III. j-j coupling and isospin dependence.  }
\author
{F.Arias de Saavedra\\ 
{\small Departamento de Fisica
Moderna, Universidad de Granada,} \\
   {\small \sl E-18071 Granada, Spain} \\ \\
     G.Co'\\
{\small Dipartimento di Fisica, Universit\`a di Lecce and} \\
{\small INFN, Sezione di Lecce,}\\
{\small \sl I-73100 Lecce, Italy }  \\ \\
     A.Fabrocini \\
{\small Dipartimento di Fisica, Universit\`a di Pisa and} \\
{\small INFN, Sezione di Pisa,}\\
{\small \sl I-56100 Pisa, Italy }  \\ \\
      S.Fantoni \\
{\small Interdisciplinary Laboratory for Advanced Studies (ILAS) and} \\ 
{\small INFN, Sezione di Trieste,}\\
{\small \sl I-34014 Trieste, Italy } } 

\date{\mbox{ }}
\maketitle
\begin{abstract}
Correlated Basis Function theory and Fermi Hypernetted Chain technique are
extended to study medium--heavy, doubly closed shell nuclei in j-j
coupling scheme,  with different single particle wave functions for
protons and neutrons  and isospin dependent two--body correlations.
Central semirealistic interactions are used. Ground state energies,
one--body densities, distribution functions and momentum distributions
are calculated for $^{12}$C, $^{16}$O, $^{40}$Ca, $^{48}$Ca and 
$^{208}$Pb nuclei. 
The values of the ground state energies provided by
isospin dependent correlations are lower than those obtained
with isospin independent correlations. 
In finite nuclear systems,
the two--body Euler equations provide  
correlation functions variationally more effective than those obtained
with the same technique in infinite nuclear matter.
\end{abstract} 
\vskip 2.0 cm
\section{Introduction}
The description of the properties of all nuclear systems, from deuteron
to nuclear matter, with a hamiltonian containing realistic two--  and
three--body potentials is one of the goals of non relativistic nuclear
many-body theory \cite{pan90}. We call realistic those two--body
potentials which reproduce the ground state properties of
the deuteron and the nucleon--nucleon scattering data. 
Three--body forces are generally required to provide a good 
description of three--body nuclei. 

The techniques to solve the Schr\"odinger equation in few--body systems 
with these potentials have reached a high degree of
sophistication and have produced excellent results \cite{che86}. 
Accurate and reliable technologies have also been developed for 
infinite nuclear systems, and, nowadays, their  
results are quite satisfactory \cite{wir88,baldo91}. 

The situation for medium and heavy nuclei is still troublesome.
The extension to these nuclei of the Monte Carlo techniques used in
few-body systems is hindered by computational difficulties related to
the  relatively high number of particles involved.  In a nucleus like
$^{40}$Ca the number of spin isospin configurations to be sampled is
of the order of the Avogadro's number. 
On the opposite side, the techniques which made possible accurate
numerical studies in infinite nuclear and neutron matters cannot be
straightfordwally applied to finite nuclear systems because they do not enjoy 
translational invariance.

In two previous papers \cite{co92,co94} we succeeded in extending
Correlated Basis Function (CBF) theory and  Fermi Hypernetted Chain
(FHNC)  cluster summation technique \cite{pan79} (successfully used to
describe nuclear matter properties) to deal with finite Fermi systems. 
In these  works the FHNC theory has been applied to study model, doubly
magic nuclei with the same number of protons and neutrons, switching off the
Coulomb interaction  and using single particle
bases in the l--s coupling scheme. 

This scheme can be adequate to describe closed shell nuclei up to
$^{40}$Ca, but it breaks down for heavier nuclei, where the 
number of neutrons  is larger than that of protons. Moreover,
while the shell closure for $A \leq 40$ corresponds to that of l--s
coupling scheme, for heavier nuclei it is necessary to use a basis
with spin-orbit splitting to reproduce the correct sequence of magic
numbers.

In the present paper we make a step forward towards a fully microscopic 
description of all doubly closed shell nuclei. We extend the FHNC 
formalism to differentiate protons and neutrons and, in addition, 
to consider single particle bases in a j--j coupling. 

In the next section we shall present the modifications 
of the FHNC equations 
needed to describe the above nuclear systems.                 

Nuclei are described as a mixture of neutrons and
protons, with possibly different populations. 
This is reflected by the presence of distribution
functions depending on the isospin of the reference nucleons. 
In a single particle basis generated by a spin--orbit term, 
the third components of the spin is no longer a good quantum number
to describe the single--particle wave functions. For this reason, 
 a new type of statistical link appears in the  
FHNC equations. 

We shall discuss the influence of the aforementioned 
FHNC modifications on 
the calculation of the nuclei ground state energy in section 3.

The results of the calculations of binding energies, density and momentum
distributions, for various doubly closed shell nuclei are presented in 
section 4. These calculations have been performed using 
central nucleon--nucleon interactions ($v_4$) with spin and isospin
dependence and without tensor components.  The correlation
functions have isospin dependence.
Since these interactions and correlations  are not yet fully realistic,
the  obtained results cannot be immediately compared with the
experimental data. These are, however, the first, true microscopic
calculations in such heavy nucleus as $^{208}$Pb and they represent 
a necessary step towards an {\sl ab initio} description of heavy nuclear
systems starting from a realistic nuclear hamiltonian.

\section{Extension of the FHNC method}

In the framework of CBF theory and along the lines of the formalism presented 
in \cite{co92}, the ground state of a nucleus with $Z$ protons and $N$ neutrons
($A=Z+N$) is described by the correlated wave--function: 
\begin{equation}
\Psi (1,\ldots,A)= F(1,\ldots,A) \Phi_{N,Z}(1,\ldots,A)
\end{equation}
where $F$ is an A--body correlation operator we shall specify later. The
function  $\Phi_{N,Z}$ is a Slater determinant   
of single particle (s.p.) wave functions $\phi_i^\alpha(i)$ 
($\alpha=p,n$) generated by the s.p. hamiltonian
\begin{equation}
h_{sp}^\alpha (i) = -\frac{\hbar^2}{2m} \nabla^2_i + U^\alpha (i) .
\end{equation}
A proper description of the sequence of magic numbers is achieved
only if the s.p. hamiltonian contains a spin--orbit
interaction. In this case, the  s.p. wave functions are
classified in terms of the total angular momentum $j$ and are 
eigenvectors of $j^2$ and $j_z$.  
We express them as:
\begin{equation} \label{spwave}
\phi^{\alpha=p,n}_{nljm} ({\bf r}_i) \ = \ R_{nlj}^\alpha
(r_i) \sum_{\mu,s} \langle l \mu \onehalf s \mid j m \rangle Y_{l \mu}
(\hat{r}_i) \chi_s (i) ,
\end{equation} 
where 
$\chi_{s}$ are the spin wavefuntions, 
$Y_{l \mu}$ the spherical harmonics and $\langle l \mu 1/2 
s \mid j m \rangle$  the Clebsch-Gordan coefficients. 

The FHNC equations can be written in terms of the one--body densities  and
the two--body distribution functions :
\begin{eqnarray}
\rho_{1}^{\alpha} ({\bf r}) & = & \langle \Psi^* \sum_{k=1}^A 
\delta ({\bf r} - {\bf r}_k) P^{\alpha}_k\Psi \rangle \\
\rho_{2,q}^{\alpha \beta} ({\bf r},{\bf r}') & = & 
\langle \Psi^*
\sum_{k \ne l=1}^A \delta ({\bf r} - {\bf r}_k)P^{\alpha}_k
\delta ({\bf r}' - {\bf r}_l) P^{\beta}_l 
O^q_{kl}\Psi \rangle~~~q=1,..4~ , 
\end{eqnarray}
where $P^{\alpha}_k$ is the projection operator on the $\alpha=p,n$ 
state of the $k$--nucleon and we have defined
\begin{equation}
\langle X \rangle = 
\frac{\int d \tau X}{\int d \tau \mid \Psi \mid^2} ,
\end{equation}
$d \tau$ meaning integration over the spatial coordinates as
well as sum over the spins. The index $q=1,..4$
labels the operatorial component of $\rho_{2,q}^{\alpha \beta}$. This 
dependence is the same as that of the $v_4$ potentials adopted,
\begin{equation}
v_4(1,2)=\sum_{q=1,4} v^{(q)}(r_{12}) O^q_{12},
\end{equation}
with $O^1_{12}=1, O^2_{12}=\nsigma_1 \cdot \nsigma_2, 
 O^3_{12}=\ntau_1 \cdot \ntau_2$ and 
$O^4_{11}=\nsigma_1 \cdot \nsigma_2 \ntau_1 \cdot \ntau_2$.

The uncorrelated density matrices are given by:
\begin{eqnarray}
\rho_0^\alpha({\bf r}_1 , {\bf r}_2) & = & \sum_{k}
\phi_k^{\alpha *} ({\bf r}_1) \phi_k^\alpha({\bf r}_2) \nonumber \\
 & = & \sum_{s,s'} \rho_0^{s s' \alpha} ({\bf r}_1 , {\bf r}_2) 
\chi_s^\dagger (1)\chi_{s'} (2) ,
\end{eqnarray}
and their spin dependent parts are:
\begin{eqnarray}
\label{UNDEN} \nonumber
\rho_0^{s s' \alpha} ({\bf r}_1 , {\bf r}_2) & = & \sum_{n,l,j} 
R_{nlj}^\alpha(r_1) R_{nlj}^\alpha(r_2) \\ 
 & & \sum_{\mu,\mu',m} \langle l \mu \onehalf s \mid j m \rangle 
\langle l \mu' \onehalf s' \mid j m \rangle Y_{l \mu}^* (\hat{r}_1)
Y_{l \mu'} (\hat{r}_2)  .
\end{eqnarray}
While in the $|l^2 s^2 l_z s_z >$ representation  
$\rho_0^{ss'\alpha}$ is diagonal in the spin variables \cite{co92},
in the $| l^2 s^2 j^2 j_z >$ one it depends on the third 
components of spin $(s,s')$. The same  dependence is also
present in the correlated density matrices, 
\begin{eqnarray} 
\label{rhoss}
\nonumber
\rho^{ss'\alpha}({\bf r}_1, {\bf r}_{1'}) & = &
\frac{A}{\langle \Psi \mid \Psi \rangle} \int d^3{r}_2 \ldots
d^3{r}_A \ \ \chi_s(1) \chi_{\alpha}(1)\\
 & & \Psi^* (1,2,\ldots,A) \Psi (1',2,\ldots,A) 
\chi^\dagger_{s'}(1') \chi^\dagger_{\alpha}(1'),
\end{eqnarray}
where a sum over all the spin coordinates is implied.

We found convenient to consider
separately the uncorrelated density (\ref{UNDEN}) for 
pairs of particles having either the same or opposite third spin component.
For this reason we define
\begin{eqnarray}
\label{rhoprima}
\rho_{0,P}^{\alpha} ({\bf r}_1 , {\bf r}_2) & = & \frac 1 {8 \pi}
\sum_{n,l,j} (2j+1) R_{nlj}^\alpha(r_1) R_{nlj}^\alpha(r_2) P_l (\cos
\theta_{12}) , \\
\label{rhoseconda}
\rho_{0,A}^{\alpha} ({\bf r}_1 , {\bf r}_2) & = & \frac 1 {4 \pi}
\sum_{n,l,j} (-1)^{j-l-1/2} R_{nlj}^\alpha(r_1) R_{nlj}^\alpha(r_2)
 \sin \theta_{12} P_l' (\cos \theta_{12}) ,
\end{eqnarray}
where $P_l(x)$ is the Legendre polynomial of l--th degree
and $P'_l(x)$ is its first derivative with respect to $x$.
The function defined in Eq.(\ref{rhoprima})  
corresponds to pairs of nucleons with parallel spins, $s=s'$ (
$\rho_{0,P}^{\alpha}=\rho_0^{\frac {1}{2}\frac {1}{2}\alpha}= 
\rho_0^{-\frac{1}{2} -\frac{1}{2} \alpha}$) .
 This is the only statistical correlation appearing in l--s coupling. 
The function defined in Eq.(\ref{rhoseconda}), 
$\rho_{0,A}^{\alpha}=\rho_0^{\frac{1}{2} -\frac{1}{2}\alpha}=
-\rho_0^{-\frac{1}{2}\frac{1}{2}\alpha}$, is instead a new statistical link
between nucleons  with antiparallel spins, 
$s=-s'$, due to the j--j coupling. 
This new function is antisymmetric under the exchange of $s$ with $s'$ 
 and of the spatial coordinates.
In all the observables whose mean value we have calculated (energy, density, 
momentum distribution), we found that the contribution of 
$\rho_{0,P}^\alpha$ is
much larger than that of $\rho_{0,A}^{\alpha}$.
As a matter of fact, if the nucleus is closed in l--s coupling (as for 
$^{16}$O and $^{40}$Ca) and no spin--orbit term is
included in the s.p. potential, the contribution of $\rho_{0,A}$
vanishes.

The uncorrelated statistical functions appear in the cluster
expansion of the two--body distribution functions and of the one--body
density matrices forming closed exchange loops. 
In l--s coupling, all the nucleons involved in the loops must have 
the same third spin component. In j--j coupling this is no longer true, 
as there may be statistical links between particles with
opposite third components. In general, the contribution from a
N--particle  exchange  loop is given by:
\[
\begin{array}{l}(-1)^{N-1}\rho_0^\alpha({\bf r}_1,{\bf r}_2)
\rho_0^\alpha({\bf r}_2,{\bf r}_3) \ldots
\rho_0^\alpha({\bf r}_N,{\bf r}_1) =
\\  \\ {\displaystyle (-1)^{N-1}\sum_{s_1,\ldots,s_N} 
\rho_0^{s_1 s_2 \alpha} ({\bf r}_1 , {\bf r}_2)
\rho_0^{s_2 s_3 \alpha} ({\bf r}_2 , {\bf r}_3) \ldots
\rho_0^{s_N s_1 \alpha} ({\bf r}_N , {\bf r}_1) } .
\end{array}
\]

\section{Energy, density and momentum distribution expectation values}

In the previous section we have discussed the modifications
induced by j--j coupling and $Z \ne N$ on the uncorrelated density matrix.
In this section we will shortly discuss how the FHNC equations change 
and how the expectation values of energy, density and 
momentum distribution are calculated. 

The structure of the FHNC equations depends on the adopted correlation 
function. We use an isospin dependent correlation, 
since we want to distinguish 
protons  from neutrons, and we choose the following form for $F$,
which allows us to consider different correlations for 
different pairs of nucleons: 
\begin{equation}
\label{corfun}
F(1,\ldots,A) = \prod_{1=i<j}^A f_{\alpha \beta}(r_{ij}) 
P^\alpha_i P^\beta_j .
\end{equation}

The form of the correlation operator (\ref{corfun}), and specifically 
the fact that the involved operators commute among themselves, 
allows us to use, in the cluster expansion, the diagrammatic rules
presented in  Ref.\cite{co92}, with only slight modifications due to the
j--j coupling scheme.  Since the system is  a mixture of different
fermions, the correlated density matrices for $pp$, $pn$ and $nn$ are
different. Furthermore, we have to consider the presence in the cc--FHNC
chains of the  new statistical link  $\rho_{0,A}^\alpha$. The new FHNC 
expressions of the two--body distribution functions and of the one--body
density matrices are given in the
Appendices A and B, respectively.

The calculation of the expectation value of the energy has been performed  
along the same lines followed in Ref. \cite{co92}. 

The evaluation of the linetic energy has been done using
the Jackson--Feenberg. This allows us for eliminating terms of the form
$(\nabla_iF)(\nabla_i\Phi_0)$, involving three--body operators expectation
values \cite{JF}. 
Following the same notation of Ref.\cite{co92}, we express 
the kinetic energy as:
\begin{equation} 
\label{t}
<T>=T_{JF}=T_F+T_\phi^{(1)}+T_\phi^{(2)}+T_\phi^{(3)}.
\end{equation}

In the previous equation we have indicated with
$T_F$ those terms where the kinetic energy operator 
acts on the correlation operator $F$ only. 
Their contribution is given by:
\begin{equation} T_F = - \frac{\hbar^2}{4m} \sum_{\alpha,\beta}
\int d^3 r_1 d^3 r_2 \rho_{2,1}^{\alpha \beta} ({\bf r}_1,{\bf
r}_2) t[f_{\alpha \beta}(r_{12})] , 
\end{equation}
where $t[f]$ has been defined as:
\begin{equation}
t[f(r)] \equiv {1\over f^2(r)}
\Bigl\{f(r)f''(r)+
{2\over r}f(r)f'(r)-f'^2(r)\Bigr\}.
\end{equation}
The remaining $T_\phi^{(n=1,2,3)}$ components are the contributions 
of those terms containing the operator 
acting  on the uncorrelated s.p. wave functions.
$T_\phi^{(1)}$ is the sum of the cluster diagrams in which the
external point 1, argument of the one--body kinetic energy operator $\hat T_1$, 
is not involved in any exchange (see Eq. 2.24 in \cite
{co92}). We obtain the expression:
\begin{equation}
T_\phi^{(1)} = - \frac{\hbar^2}{4m} \sum_{\alpha} \int d^3 r_1
\rho_{T1}^\alpha ({\bf r}_1) \xi_e^\alpha ({\bf r}_1) , 
\end{equation}
where $\rho_{T1}^\alpha$ is defined as:
\begin{equation}
\rho_{T1}^\alpha({\bf r}_1)=
\sum_i\phi^{\alpha *}_i({\bf r}_1)\nabla^2_1
\phi^\alpha_i({\bf r}_1)
-\sum_i \nabla_1\phi^{\alpha *}_i({\bf r}_1)\cdot
\nabla_1\phi_i^\alpha({\bf r}_1),
\end {equation}
and $\xi_e^\alpha$ is defined in appendix A.

The other two parts ($T_\phi^{(2,3)}$) contain diagrams where 
the point 1 
belongs to a 2--body ($T_\phi^{(2)}$) or to a $n(>2)$--body ($T_\phi^{(3)}$) 
exchange loop. We find for $T_\phi^{(2)}$ the expression:
\begin{equation}
T_\phi^{(2)} =  \frac{\hbar^2}{4m} \sum_{\alpha} \int d^3 r_1
d^3 r_2 \xi_e^\alpha ({\bf r}_1) 
\rho_{T2}^\alpha ({\bf r}_1,{\bf r}_2)
\left[ g_{dd}^{\alpha \alpha}({\bf r}_1,{\bf r}_2) 
\xi_e^\alpha ({\bf r}_2) -1 \right] ,
\end{equation}
where 
\begin{eqnarray}
\nonumber
\rho_{T2}^\alpha ({\bf r}_1,{\bf r}_2) & = &
\rho_{0,P}^\alpha ({\bf r}_1,{\bf r}_2) \nabla_1^2
\rho_{0,P}^\alpha ({\bf r}_1,{\bf r}_2)-
\nabla_1 \rho_{0,P}^\alpha ({\bf r}_1,{\bf r}_2) \cdot
\nabla_1 \rho_{0,P}^\alpha ({\bf r}_1,{\bf r}_2) + \\
& & \rho_{0,A}^\alpha ({\bf r}_1,{\bf r}_2) \nabla_1^2
\rho_{0,A}^\alpha ({\bf r}_1,{\bf r}_2)-
\nabla_1 \rho_{0,A}^\alpha ({\bf r}_1,{\bf r}_2) \cdot
\nabla_1 \rho_{0,A}^\alpha ({\bf r}_1,{\bf r}_2) . 
\end{eqnarray}
All the FHNC quantities introduced in the above equations are
defined in appendix C.

The evaluation of  $T_\phi^{(3)}$ requires the knowledge of the three--body
distribution functions, but its leading term ($T_\phi^{(3,2)}$) 
is a function of two--body dressed FHNC quantities \cite{co92}. 
$T_\phi^{(3,2)}$ is given by:
\begin{equation}
T_\phi^{(3,2)} 
= -\frac{\hbar^2}{2m}  \sum_{\alpha} \int d^3 r_1
d^3 r_2 \left[ \rho_{T3,P}^\alpha ({\bf r}_1,{\bf r}_2)
H_{cc,P}^\alpha ({\bf r}_1,{\bf r}_2) +
\rho_{T3,A}^\alpha ({\bf r}_1,{\bf r}_2)
H_{cc,A}^\alpha ({\bf r}_1,{\bf r}_2) \right] , 
\end{equation}
where
\begin{eqnarray} 
\nonumber
H_{cc,X}^\alpha ({\bf r}_1,{\bf r}_2) & = & \xi_e^\alpha (r_1)
\{   \xi_e^\alpha (r_2) [ (g_{dd}^{\alpha \alpha}
 ({\bf r}_1,{\bf r}_2) -1 ) 
N_{cc,X}^\alpha ({\bf r}_1,{\bf r}_2)
+N_{cc,X}^{(x)\alpha} ({\bf r}_1,{\bf r}_2) ] \\
& + & 
(\xi_e^\alpha (r_2) -1) N_{cc,X}^{(\rho)\alpha} ({\bf r}_1,{\bf r}_2)
+  \xi_e^\alpha (r_2) g_{dd}^{\alpha \alpha}
({\bf r}_1,{\bf r}_2) E_{cc,X}^\alpha ({\bf r}_1,{\bf r}_2)
\; \},
\end{eqnarray}
with $X=P,A$. The remaining term $T_\phi^{(3,3)}$ contains a 
three-body operator and it is known to be negligible, 
both in nuclear matter and N=Z nuclei 
\cite{co92}. Therefore, it has been  neglected.

The center of mass kinetic energy, $T_{cm}$,  has to be subtracted 
from $<H>$ to get the energy mean value $E$. $T_{cm}$ is given by:

\begin{equation}
\label{tcm}
T_{cm} = - \frac{\hbar^2}{4mA} \sum_{\alpha=p,n} \int d^3 r_1 \left(
\rho_{T1}^\alpha({\bf r}_1) - \int d^3 r_2
\rho_{T4}^\alpha({\bf r}_1,{\bf r}_2) \right) .
\end{equation}

The expressions of  the $\rho_{Tx}^\alpha$  functions are given in 
appendix C.

The potential energy $<V>$ may be written as:
\begin{equation}
\label {v}
<V> = \frac {1}{2}  \sum_{\alpha,\beta=p,n}
\int d^3 r_1  d^3 r_2 
\sum_{q=1,4} v^{(q)}(r_{12}) \rho_{2,q}^{\alpha \beta}({\bf r}_1,{\bf r}_2) .
\end{equation}

The two--body distribution functions can be decomposed into 
a direct part, summing those contributions where the particles $1$ 
and $2$ do no belong to the same exchange loop, 
and two exchange parts, containing diagrams where the two nucleons are
involved in   the same loop with identical ($\rho_{exc,P}$) or opposite  
($\rho_{exc,A}$) third spin components:
\begin{equation}
\label{rho2q}
\rho_{2,q}^{\alpha \beta} ({\bf r}_1,{\bf r}_2) =
 a_{q}^{\alpha \beta} \rho_{dir}^{\alpha \beta} ({\bf r}_1,{\bf r}_2) + 
 b_{q}^{ \alpha \beta} \rho_{exc,P}^{\alpha \beta} ({\bf r}_1,{\bf r}_2) +
 c_{q}^{ \alpha \beta} \rho_{exc,A}^{\alpha \beta}({\bf r}_1,{\bf r}_2) .
\end{equation}
The $a,b$ and $c$ factors are given in Tab.1. The expressions
 for $\rho_{dir}$ and  $\rho_{exc,X}$ are:
\begin{eqnarray}
\nonumber
\rho_{dir}^{\alpha \beta} ({\bf r}_1,{\bf r}_2) & = &  
\xi_d^\alpha({\bf r}_1)
(\xi_d^\beta({\bf r}_2) g_{dd}^{\alpha \beta}({\bf r}_1,{\bf r}_2)+
 \xi_e^\beta({\bf r}_2) g_{de}^{\alpha \beta}({\bf r}_1,{\bf r}_2))+ \\
& & \xi_e^\alpha({\bf r}_1)\xi_d^\beta({\bf r}_2)
g_{ed}^{\alpha \beta}({\bf r}_1,{\bf r}_2) + \nonumber \\
 & & \xi_e^\alpha({\bf r}_1)\xi_e^\beta({\bf r}_2)
g_{dd}^{\alpha \beta}({\bf r}_1,{\bf r}_2)
\Biggl[N_{ee}^{\alpha \beta}({\bf r}_1,{\bf r}_2)+
E_{ee,dir}^{\alpha \beta}({\bf r}_1,{\bf r}_2)+
\nonumber \\
 & & (N_{de}^{\alpha \beta}({\bf r}_1,{\bf r}_2)
+E_{de}^{\alpha \beta}({\bf r}_1,{\bf r}_2))
(N_{ed}^{\alpha \beta}({\bf r}_1,{\bf r}_2)
+E_{ed}^{\alpha \beta}({\bf r}_1,{\bf r}_2)) \Biggr] , 
 \\
\nonumber
\rho_{exc,X}^{\alpha \beta} ({\bf r}_1,{\bf r}_2) & = & 
 \xi_e^\alpha({\bf r}_1)\xi_e^\beta({\bf r}_2)
g_{dd}^{\alpha \beta}({\bf r}_1,{\bf r}_2)
\Biggl[E_{ee,exc,X}^{\alpha \beta}({\bf r}_1,{\bf r}_2)- \\
 & & 2(N_{cc,X}^{\alpha}({\bf r}_1,{\bf r}_2)+ 
 E_{cc,X}^{\alpha}({\bf r}_1,{\bf r}_2) - 
 \rho_{0,X}^\alpha ({\bf r}_1,{\bf r}_2)) \nonumber \\
 & &  (N_{cc,X}^{\beta}({\bf r}_1,{\bf r}_2)
+E_{cc,X}^{\beta}({\bf r}_1,{\bf r}_2)-
\rho_{0,X}^\beta ({\bf r}_1,{\bf r}_2)) \Biggr]  . 
\end{eqnarray}

We consider also the contribution
of the Coulomb interaction acting between the protons. Its
expectation value is:
\begin{equation}
\label{vc}
<V_c> = \frac 1 2 \int d^3 r_1d^3 r_2
\frac {e^2}{r_{12}} \rho_{2,1}^{pp}({\bf r}_1,{\bf r}_2) . 
\end{equation}

The mean value of the energy is then given by $E=T_{JF}+<V>+<V_c>-T_{cm}$.

Since the one--body density matrices depend on the third components of
the spin (Eq.(\ref{rhoss})), also the momentum distributions have this 
 dependence :
\begin{equation}
n^{s s' \alpha}(k) = \frac 1 A \int d^3 r_1 d^3 r_{1'}
e^{i{\bf k} \cdot ({\bf r}_1-{\bf r}_{1'})} \rho^{s s' \alpha}
({\bf r}_1 , {\bf r}_{1'}).
\end{equation}
The spin averaged momentum distribution, summed over  all the possible spin 
components, is:
\begin{equation}
\label{nk}
n^{\alpha}(k) = \sum_{s,s'} n^{s s' \alpha}(k),
\end{equation}
 and, in the following, we shall always refer to the momentum distribution
as defined by Eq.(\ref{nk}). The FHNC derivation of the one--body density
matrices are given in appendix B.

 We conclude this section by listing some sum rules (SR) that must be satisfied 
by the densities, the two--body distribution functions and the momentum 
distributions.  
They are particularly relevant since provide information 
about the accuracy of the approximations used to solve the FHNC equations.

 For the densities and the distribution functions, the following SR's hold:

\begin{eqnarray}
\label{sump}
S_p & = & \frac{1}{Z}\int d^3 r \rho^p_1 ({\bf r})  =  1  , \\
\label{sumn}
S_n & = & \frac{1}{N}\int d^3 r \rho^n_1 ({\bf r})  =  1  , \\
\label{sumpp}
S_{pp} & = & \frac{1}{Z(Z-1)}\int d^3 r_1 d^3 r_2 
              \rho_{2,1}^{pp} ({\bf r}_1,{\bf r}_2)  = 1 , \\
\label{sumpn}
S_{np(pn)} & = & \frac{1}{ZN}\int d^3 r_1 d^3 r_2 
        \rho_{2,1}^{np(pn)} ({\bf r}_1,{\bf r}_2) 
             = 1 , \\
\label{sumnn}
S_{nn} & = & \frac{1}{N(N-1)} \int d^3 r_1  d^3 r_2 
        \rho_{2,1}^{nn} ({\bf r}_1,{\bf r}_2) = 1 . 
\end{eqnarray}
In nuclei, where the levels with the same value
of the orbital angular momentum are saturated (as $^{16}$O and $^{40}$Ca),
the spin--SR $S_{\sigma}$,

\begin{equation}
\label{sSR}
S_{\sigma}  =  \frac{1}{3A}\int d^3{r}_1 d^3{r} \,\,
            [ \rho_{2,2}^{pp} ({\bf r}_1,{\bf r}_2) +  
              \rho_{2,2}^{nn} ({\bf r}_1,{\bf r}_2) ]  ,
\end{equation}

has to satisfy the condition $S_\sigma=-1$. This is no longer true for
the nuclei $^{12}$C, $^{48}$Ca and $^{208}$Pb. 
However, as the correlations do not contain spin flip terms, $S_\sigma$ in the
correlated nucleus must have the same value as in the uncorrelated one
($S_\sigma^{corr}=S_\sigma^{unc}$).

The momentum distributions must obey the zeroth momentum SR: 
\begin{eqnarray}
\label{sumnk}
 MD_0^{\alpha}=\frac{1}{(2 \pi)^3} \int d^3 k n^\alpha(k)  = x_\alpha ,
 \end{eqnarray}
 where $x_\alpha$ is the proton ($x_p=Z/A$) or neutron fraction 
($x_n=N/A$), and the second momentum SR: 
\begin{equation}
\label{sumknk}
  MD_2=\frac{ \hbar^2}{2 m (2 \pi)^3} \int d^3 k k^2 [Zn^p(k)+Nn^n(k)] 
  = <T> . 
\end{equation}

\section{Results}

We have applied the theoretical framework presented in the previous sections 
to the study of the ground state properties of the 
$^{12}$C, $^{16}$O, $^{40}$Ca, $^{48}$Ca and $^{208}$Pb nuclei. 

Our calculations have been done using two different kinds of correlation
 functions. One has a gaussian  behaviour:
\begin{equation}
\label{gauss}
f^G_{\alpha\beta}(r)= 1 - A_{\alpha\beta} \exp (-B_{\alpha\beta} r^2),
\end{equation}
where $A_{\alpha\beta}$ and $B_{\alpha\beta}$ are taken as 
variational parameters, fixed by minimizing the FHNC energy. 

The second  correlation function is obtained by the minimization of 
the energy at the second order of the cluster expansion, $<H_2>$, with 

\begin{eqnarray}
<H_2> & = & -\frac{\hbar^2}{4m} \sum_{\alpha=p,n} \int d^3 r_1
\left( \rho_{T1}^\alpha({\bf r}_1) + \int d{\bf r}_2
\rho_{T2}^\alpha ({\bf r}_1,{\bf r}_2) \right) + \\
& & \hspace{-2.3cm} \sum_{\alpha,\beta=p,n} 
\int d^3 r_1 d^3 r_2 \left[
Q_{\alpha \beta} ({\bf r}_1,{\bf r}_2) 
f_{\alpha \beta}^2 (r_{12})- 
P_{\alpha \beta} ({\bf r}_1,{\bf r}_2) 
\left(f_{\alpha \beta} (r_{12}) \nabla_{12}^2 f_{\alpha \beta} (r_{12})
-(f_{\alpha \beta}^\prime (r_{12}))^2 \right) \right] . \nonumber
\end{eqnarray}
 
The solution of  $\delta <H_2> / \delta f_{\alpha\beta} = 0 $   
 provides the Euler Correlations $f^E_{\alpha\beta}$.

The quantities $Q(P)_{\alpha \beta}$ are generalizations to the isospin 
dependent case of those of Ref.\cite{co92}:

\begin{eqnarray}
Q_{\alpha \beta} ({\bf r}_1,{\bf r}_2) & = & \delta_{\alpha\beta}
\frac{\hbar^2}{4m}\rho_{T2}^\alpha ({\bf r}_1,{\bf r}_2)
+\frac 1 2 \rho_{0}^\alpha ({\bf r}_1)\rho_{0}^\beta ({\bf r}_2)
V_{dir}^{\alpha \beta}(r_{12})- \\
& & \left[
\rho_{0,P}^\alpha ({\bf r}_1,{\bf r}_2)
\rho_{0,P}^\beta ({\bf r}_1,{\bf r}_2)
V_{exc,P}^{\alpha \beta}(r_{12})+
\rho_{0,A}^\alpha ({\bf r}_1,{\bf r}_2)
\rho_{0,A}^\beta ({\bf r}_1,{\bf r}_2)
V_{exc,A}^{\alpha \beta}(r_{12}) \right] , \nonumber \\
P_{\alpha \beta} ({\bf r}_1,{\bf r}_2) & = & 
\frac{\hbar^2}{4m} \left\{ \rho_0^\alpha ({\bf r}_1)
\rho_0^\beta ({\bf r}_2)-2 \delta_{\alpha\beta}\left[
(\rho_{0,P}^\alpha ({\bf r}_1,{\bf r}_2))^2+
(\rho_{0,A}^\alpha ({\bf r}_1,{\bf r}_2))^2
\right] \right\} ,
\end{eqnarray}

where:

\begin{eqnarray}
V_{dir}^{\alpha \beta}(r_{12}) & = &
\delta_{\alpha\beta}[v^{(1)}(r_{12})+v^{(3)}(r_{12})]+
(1-\delta_{\alpha\beta})[v^{(1)}(r_{12})-v^{(3)}(r_{12})],  \\
V_{exc,P}^{\alpha \beta}(r_{12}) & = & 
  -2\delta_{\alpha\beta}[v^{(1)}(r_{12})+v^{(3)}(r_{12})+
3v^{(2)}(r_{12})+3v^{(4)}(r_{12})]- \\
& &4(1-\delta_{\alpha\beta})[v^{(3)}(r_{12})+3v^{(4)}(r_{12})], \nonumber \\
V_{exc,A}^{\alpha \beta}(r_{12}) & = &  
  -2\delta_{\alpha\beta}[v^{(1)}(r_{12})+v^{(3)}(r_{12})-
v^{(2)}(r_{12})-v^{(4)}(r_{12})]- \\
& &4(1-\delta_{\alpha\beta})[v^{(3)}(r_{12})-v^{(4)}(r_{12})]. \nonumber
\end{eqnarray}

 The contributions to $<H_2>$ decouple in the ($\alpha \beta$) channels, 
 so the minimizations are independently performed for each component. 
 The corresponding Euler--Lagrange equations are:

\begin{equation}
\label {eu1}
u_{\alpha \beta}^{\prime \prime} (r) - \overline{V}_{\alpha \beta}(r) 
u_{\alpha \beta} (r) = \lambda_{\alpha \beta} u_{\alpha \beta} (r) , 
\end{equation}

with:

\begin{eqnarray}
u_{\alpha \beta} (r) & = & r \sqrt{\overline{P}_{\alpha \beta} (r)} 
f_{\alpha \beta} (r) \\
\overline{V}_{\alpha \beta} (r) ,  & = &
\frac 1 {4 \overline{P}_{\alpha \beta} (r)} \left[
\nabla^2 \overline{P}_{\alpha \beta} (r) +
2 \overline{Q}_{\alpha \beta} (r)+
\frac{\overline{P}_{\alpha \beta}^\prime (r)^2}
{\overline{P}_{\alpha \beta} (r)} \right]  .
\end{eqnarray}

The
$\overline{P}_{\alpha \beta}$ and  $\overline{Q}_{\alpha \beta}$ 
terms are obtained
by integrating $P_{\alpha \beta} ({\bf r}_1,{\bf r}_2)$ and
$Q_{\alpha \beta} ({\bf r}_1,{\bf r}_2) $ over ${\bf r}_1$ and
${\bf r}_2$ and keeping $r_{12}$ fixed: 
\begin{equation}
\overline{X} (r_{12}) = \frac 1 {r_{12}} \int_0^\infty dr_1 r_1
\int_{|r_{12}-r_1|}^{r_{12}+r_1} dr_2 r_2 X(r_1,r_2,r_{12}) , 
\end{equation}
where $X$ can be either $P$ or $Q$.

Eqs. (\ref{eu1}) are solved under the constraints: 
 
\begin{equation}
f_{\alpha \beta} (r \ge d_{\alpha \beta}) =1 \hspace{2.0cm}
f_{\alpha \beta}^\prime (r \ge d_{\alpha \beta}) =0 ,
\end{equation}

where the healing distances $d_{\alpha \beta}$ are variational parameters. 
The same value for all the healing distances ($d_{\alpha \beta}=d$)
has been used in this paper. 

In some calculations, we have also adopted  the Average
Correlation Approximation (ACA), consisting in using an unique correlation,
independent on the isospin of the nucleons,
for both types of correlations, either gaussian or Euler.

For the gaussian correlation, ACA simply consists in using the same values of
the $A$ and $B$  constants for each channel.

The ACA Euler equation is immediately obtained by summing over the
$\alpha,\beta$ effective potentials. The resulting equation is:
\begin{equation}
\label {eu2}
u^{\prime \prime} (r) - \overline{V}(r) u(r) = \lambda u (r) , 
\end{equation}
where:
\begin{eqnarray}
u(r) & = & r \sqrt{\overline{P}(r)} f(r) , \\
\overline{V}(r) & = &
\frac 1 {4 \overline{P} (r)} \left[ \nabla^2 \overline{P} (r) +
2 \overline{Q}(r)+\frac{\overline{P}^\prime (r)^2}
{\overline{P} (r)} \right] , \\
\overline{P} (r) & = & \sum_{\alpha \beta} \overline{P}_{\alpha
\beta} (r) ,\\
\overline{Q} (r) & = & \sum_{\alpha \beta} \overline{Q}_{\alpha
\beta} (r) .
\end{eqnarray}

The second ingredient of the calculations is the set of
s.p. wave functions. In this paper, they 
have always been generated by a mean field potential of
Woods--Saxon type: 
\begin{equation}
\label{wood}
V_{WS}({\bf r})={-V_0 \over 1+e^{(r-R)/a} }+
\left( { \hbar \over m_\pi c} \right)^2 {1 \over r} 
{d \over dr} \left( {-V_{ls} \over 1+e^{(r-R)/a} } \right) \;
{\bf l} \cdot  {\mbox{\boldmath $\sigma$}} + V_{Coul},
\end{equation}
where $m_\pi$ is the pion mass.

As a test of accuracy of the calculations, we show in Tab.2 
the level of exhaustion of the previously discussed densities and 
distribution functions SR's. They have been calculated in the FHNC/0 
approximation, consisting in neglecting the elementary (or bridge) 
diagrams. These diagrams, designed as $E_{xy}^{\alpha\beta}$ in 
Appendix A, cannot be summed in a closed way by FHNC. Therefore, 
they are usually neglected. In some cases, their contribution is 
approximated by considering only a few low order terms.
The seemingly crude FHNC/0 approximation has been shown to be
accurate in relatively low density systems as nuclei and nuclear matter
\cite{wir88}, whereas the elementary diagrams become increasingly important 
in high density systems as liquid atomic Helium \cite{arias}. 
On the other side, it has already been pointed out that the SR's are 
evaluated just to ascertain the degree of accuracy of the approximations
employed in solving the FHNC equations. In the following, we shall always
use FHNC/0, unless differently stated.

The results shown in Tab.2 have  been obtained with the parameters of the 
ACA calculations of Tab.5, whose details will be discussed later. 
In the non ACA case, the accuracy is substantially the same.   The
SR's are very well satisfied for all the medium--heavy nuclei. The
worse situation is met in $^{208}$Pb, but the error remains less
than $2\%$.

In the last two rows of the Table we
show the ratios between the values of the spin SR of Eq.(\ref{sSR}) 
obtained in the correlated system and that obtained in the uncorrelated one
($S_\sigma^{corr}/S_\sigma^{unc}$).
In the row labelled $S_{\sigma,0}$ the correlated value has been evaluated 
in FHNC/0, while in the row labelled $S_{\sigma,1}$ the contribution of the
first order exchange elementary diagram has been included (see Ref.\cite{co92}
 for a more extensive discussion of this point). This last approximation has
been termed as  FHNC--1. 
In agreement with the findings of Ref. \cite{co92}, Tab.2 shows that 
the FHNC--1 approximation largely improves the accuracy of the 
calculated spin sum rules with respect to FHNC/0.

A first set of calculations has been performed to
investigate the relevance of the  j--j coupling, 
of the  Coulomb interaction and of the separated
treatment of protons and neutrons. In order to  have a better control on
their contributions, we have used
the same set of Woods-Saxon parameters, identical  for protons and
neutrons  and without spin--orbit term. 
In addition, we have taken the same s.p. potentials in $^{12}$C and 
$^{16}$O and for the Ca isotopes, as shown in Tab.3, to investigate 
the influence of the unsaturated l shells. The semi--realistic S3
interaction of Afnan and Tang \cite{afn68}, supplemented in the odd
channels with the repulsive interaction given by the repulsive terms of
the even channels as discussed in Ref. \cite{gua81}, has been adopted in 
conjunction with the Euler ACA correlation function.

In Tab.4 we present the  results of the ground state energies. 
Column F1 has been obtained by switching off the Coulomb interaction and the
statistical correlation arising from the j--j coupling, Eq.(\ref{rhoseconda}). 
Its effect can be seen by comparing the results of column F1 with those 
of column F2, where the j--j coupling has been reinstated. 
Only  $^{12}$C, $^{48}$Ca and $^{208}$Pb nuclei, which have some unsaturated l 
shell, are affected by this correlation and its influence turns out to 
be rather small.

The F3 column shows the results obtained including the Coulomb
interaction in the two--body hamiltonian. 
The contribution of the nuclear interaction
($V$ in the Table) to the binding energy is about the same for 
$^{40}$Ca,  $^{48}$Ca and  $^{208}$Pb. These nuclei are large enough  to
allow the nuclear interaction to saturate. The
contribution of the Coulomb interaction $V_c$, because of the infinite range
of the force, increases like the number of proton pairs, as expected.

The results of the F4 column have been obtained by inserting the Coulomb potential
also in the mean field. We have used the potential generated by an uniform
spherical charge distribution. Other choices did not make any difference
from the numerical point of view.

A fully implemented variational principle would require performing the 
minimization of the energy functional respect to both the s.p.
wave functions and the correlation function parameters. 
This implies a large parameter space minimization. Because the
requested numerical effort is heavy, and also because we are not yet
considering fully realistic interactions, 
the set of single particle wave functions has been kept fixed and the
minimization has been done only respect to the correlation function. 

 Moreover, for all the columns of Tab.4 we have used the same 
 correlations 
(obtained by minimizing the F1 energies). Therefore, we cannot draw 
any definite conclusion on whether the l--s coupling is or is not 
variationally preferred with respect to the j--j coupling, except that 
we find very small differences in the two cases. 
 Similarly, we should not be surprised by the fact that the 
 inclusion of the Coulomb effects in the wave function (column F4) 
 does not seem to lower the energies with respect to their perturbative 
 estimates of column F3. 

 In all the calculations we shall discuss henceforth
the s.p. wave functions have been generated by Wood-Saxon
potentials whose parameters, given in Tab.5, are taken from literature
\cite{rin78}, and have been fixed to reproduce the s.p. energies
around the Fermi surface and the rms charge radii. 

Table 6 gives the results obtained with the S3 interaction and various 
correlation functions. 
The $EU$ rows show  the results with the isospin 
dependent Euler correlation function, the ACA rows those with the ACA 
Euler correlation  and the $G$ rows those with the gaussian ACA
correlation.  The columns present the contributions of the various terms 
building up the average value of the binding energy. The binding energies 
per nucleon, $E/A$, have been calculated by subtracting the center of
mass energy. In the last column the healing distances, for the
minimum of the energy with the Euler correlations, are given. 
Note that the EU calculations have been performed using the same value
of the healing distance for all the isospin channels.

For all the nuclei we have considered, the isospin dependent correlation 
function results to be variationally preferred respect to the simpler 
ACA functions. This is mainly produced by a large decrease  of the 
potential energy, which overcompensates the increase of the 
kinetic energy. The poorest choice is the ACA gaussian one.

The correlation functions used in the ACA Euler and gaussian calculations 
of Tab.6 are shown in Fig.1. They are very similar in all the nuclei we
 have studied and all the finite system Euler correlations overshoot one.

Also in the isospin dependent case, where $f^{pp}$, $f^{nn}$ and $f^{np}$
 are different even if we  assume the same healing distance
for all of them, the correlations do not show a strong dependence on the
type of nucleus considered. This is due to the fact that the
correlations are mostly sensitive to the short range behaviour of the
nuclear interaction. 
Fig.2 compares the isospin dependent Euler correlations with the ACA
Euler ones for $^{12}$C and $^{208}$Pb. $f^{pp}$ and $f^{nn}$ are close to 
$f^{ACA}$, but $f^{np}$ shows large differences. This rich isospin
structure is responsible for the better variational behaviour of the
correlation. 

It is interesting to compare the finite nuclei Euler
correlations with those of infinite nuclear matter. In this case,  
 we put $f^{nn}=f^{pp}=f_{NM}^{T=1}$ and 
 $f^{np}=\frac{1}{4}[3f_{NM}^{T=1}+f_{NM}^{T=0}]$, where $T=0,1$ is 
the total isospin. The nuclear matter correlations are always smaller
than one, therefore the overshooting feature of the Euler correlations in
the nuclei seems to be ascribed to the finite nature of the systems. 
Nuclear matter correlations have been often used  in finite 
nuclei \cite{SteveO16} on the assumption that their  
short range behaviour is little affected by 
the size of the system. However, we find not negligible differences 
 and improved energies by using correlations generated by the finite 
 system Euler equations. For instance, in $^{12}$C 
$f_{NM}^T$ gives $E/A=-2.12 MeV$ at $d=2.0 fm$, to be compared with 
the $E/A=-4.65 MeV$ value of the Euler correlation.

The proton one--body densities, $\rho_1^p({\bf r})$, 
calculated with the Euler
and gaussian correlations are given in Fig.3 for $^{12}$C, $^{16}$O and the Calcium isotopes and Fig.5 for $^{208}$Pb,
 and compared with their uncorrelated counterparts, $\rho_0^p({\bf r})$,
(full lines). The dotted curves 
have been obtained with the gaussian--ACA correlations and all of them
are lower than the uncorrelated densities at low $r$--values. 
On the other hand, those obtained in the Euler (dashed--dotted lines) 
and ACA Euler (dashed lines) calculations are larger than 
$\rho_0^p({\bf r})$.
This indicates that the correlation effect on $\rho_1^p({\bf r})$ at small
distances may strongly depend on the type of correlation employed.
The results of the figures have been obtained for those values
of the variational parameters giving the minimum energy. 
We found correlated densities smaller than $\rho_0^p({\bf r})$ 
in the nuclear center when we used values of the healing distance far
from those producing the energy minima.

Similar results have been obtained for the neutrons density
distributions.

In order to check the validity of our approximations  and  
their sensitivity to the choice of the
nucleon--nucleon interaction, we have  performed ACA Euler 
calculations with the Brink and Boeker B1 central potential\cite{bri67}. 
The B1 potential is not a microscopic one, as it does not fit 
two--body data. It is an effective interaction which reproduces 
the nuclear matter and $^4$He binding energies in Hartree-Fock theory. 
This potential has been used in Ref.\cite{co92}
to compare its FHNC results with the available, exact Variational 
Monte Carlo calculations in $^{16}$O. 

The accuracy of the  SR's for the B1 potential is the same as for S3.
The ground state expectation values of the various terms of the
hamiltonian are shown in Tab. 7. 
 Also in B1 case, the  correlation functions, shown in Fig.6, are very
similar for all the nuclei considered. However, some 
differences from those obtained with the S3 interaction are present. 
The B1 correlations have a value of  about $0.55$ at $r=0$ fm, 
versus the $0.37$ value of the S3 case. In
addition, the overshooting is rather small.

The differences in the correlation functions have some consequences
on the proton density distributions, shown in Figs.3 and 5
by the dashed--doubly--dotted lines. All these
distributions are smaller than the uncorrelated ones, in contrast with
the ACA Euler results with S3.

We have studied the effect on the energy of the insertion of the dressed, 
lowest order exchange elementary diagrams in FHNC--1. 
The results are summarized in Tab.8, where the FHNC-1 energies per nucleon 
(columns $E_1$) obtained with the
Euler--ACA correlations for both S3 and B1 interactions are given. 
In addition we give the percentile deviations $\Delta$ from the FHNC/0
energies:
\begin{equation}
\label{delta}
\Delta= 100 \frac {E_0-E_1} {E_0}
\end{equation}
where $E_1$ and $E_0$ are the FHNC--1 and FHNC/0 energies respectively.

Confirming the finding of Ref.\cite{co92}, $E_1$ is always higher than 
$E_0$. Moreover, the influence of this class of  elementary
diagram becomes  small in large nuclei, in agreement with nuclear matter
results \cite{wirpriv}.

The proton MDs, $n^p(k)$, obtained for S3 are shown in Fig.4 for 
$^{12}$C,$^{16}$O,$^{40}$Ca and $^{48}$Ca
nuclei, and, in the lower panel of Fig.5 for $^{208}$Pb. 
In the same figure we show the 
uncorrelated distributions (full lines) and $n^p(k)$ for the B1 model with 
the ACA Euler correlations. The MD sum rules of Eqs.(\ref{sumnk},\ref{sumknk}) 
are given in Tab.(9) for different wave functions. The $MD^\alpha_0$ sum rules 
are very well satisfied, with errors less than 1\%, except one single
case.  The  $MD_2$ sum rules are also well satisfied, even if with a
lower accuracy.  However, for the best variational case (the Euler
correlation) the error in  $MD_2$ is, at most, about 5\%.

 All the FHNC results show the same known features: the correlated
results  are orders of magnitude larger that the uncorrelated ones at
high momentum  values. In particular, the isospin dependent correlations
enhance the tails of the MDs  more than the Jastrow, isospin independent
choice. This result is in line  with the findings of
Ref.\cite{co94}, where the effect of state  dependent correlations
(having spin, isospin and tensor components) was  estimated, from CBF
Nuclear Matter calculations, in a suitable Local  Density Approximation
and compared  with a Jastrow finite nuclei FHNC  evaluation of $n(k)$ in
N=Z nuclei. State dependence was found to  increase the tails of the MDs
of about one order of magnitude more than  the Jastrow correlations. The
enhancement found here is somewhat smaller.  However, a consistent FHNC
approach, to deal with nuclear matter type  correlations, is called for
before a quantitative an definite answer  may be given.

\section{Conclusions}

This work represents a further step towards a fully microscopical
description of medium--heavy nuclei. For the first time we have
performed FHNC calculations for medium and heavy finite nuclear
systems.   The calculations have been done using 
different s.p. wave functions for protons and neutrons, in j--j
coupling, isospin dependent correlations and semirealistic, central
potentials.  The Coulomb interaction has been also considered.

The extension of the FHNC equations of Refs.\cite{co92,co94} is not 
straightforward because, due to the j--j coupling scheme,
a new type of statistical link appears.

We have used the central S3, Afnan and Tang, and B1, Brink and Boeker, 
interactions. The s.p. wave functions have been 
generated by Woods--Saxon potentials, whose parameters have been taken 
from literature.

A detailed analysis of the influence of the j--j statistical
correlations on the FHNC equations has been presented. 
Their contribution is rather small,
while the effect of the Coulomb interaction, especially for heavy nuclei, 
turns out to be relevant.

Both isospin dependent and independent correlation functions have been 
used. The first ones allow us to
distinguish between $pp$ $nn$ and $np$ channels. 
In all the nuclei considered we have found that
the $pp$ and $nn$ correlations are
quite close to each other, while the $np$ correlation shows pronounced
differences. This marked isospin structure appears to be variationally
preferred, as it largely lowers the ground state energies with
respect to the results obtained with isospin independent correlations.

At the variational minima, the correlation functions do not strongly
depend on the considered nucleus. However, they show an  
intermediate range structure, not present in the optimal FHNC
correlations  of infinite, symmetric nuclear matter. 
This difference has large effects on the energies and the
density  distributions.

The accuracy of the FHNC/0  and FHNC--1 approximations has been studied 
by examining the density and distribution functions sum rules. FHNC/0 is
found rather accurate, the largest error being less than $2\%$ in the
scalar sum rules. The inclusion of the lowest order, exchange elementary 
diagram, in FHNC--1, lowers the error in the spin sum rules to less than 
$1\%$. The FHNC--1 correction is also relevant in the ground state
energy, because the adopted potentials have large Majorana parts \cite{co92}. 
However, its contribution is negligible for large  mass nuclei,
as $^{208}$Pb. 

As far as the matter density is concerned, it appears that, given a
chosen s.p. structure, the correlations may largely and differently 
modify the density distribution. 
The results we have presented are however related to energy minima
obtained by performing variations of the correlation functions only.
Before drawing any sensible conclusion one should include tha
correlations in a fully consistent scheme. In our case, thise means to
perform the energy minimization by variating also the mean field
parameters.

The momentum distributions have been calculated within the FHNC/0 scheme
and the sum rules analysis shows that its accuracy is actually very
satisfactory.  Short range correlations produce high momentum tails,
absent in an independent  particle model. The momentum distributions
calculated with state dependent correlations show tails higher than those
produced by Jastrow, state dependent correlations.

In this paper, we have shown,  that it is possible to treat accurately 
doubly closed shell nuclei in the  j--j scheme using the FHNC cluster  
summation technique. This makes feasible the microscopical investigation of
heavy nuclei, such as $^{208}$Pb. Isospin state dependence in the
correlation  can be safely handled and provides a better variational
choice than  Jastrow correlated wave functions. This encourages us to
pursue  the goal of a microscopic description of finite nuclei with even  
more realistic correlation functions and interactions containing, 
in particular, tensor components.

\vskip 2.cm
{\bf Aknowledgements}
This project has been partially supported by the CYCIT-INFN agreement and by
the Junta de Andalucia. One of the authors (A.F.) wants to thank the
warm hospitality of CEBAF, where part of the work has been done.

\newpage
\section*{Appendix A}

This appendix contains the two--body distribution function and 
one--body density FHNC equations for the j--j coupling and 
isospin dependence. The topological 
classification of the cluster diagrams (Nodal, Composite and Elementary, 
$dd$, $de$, $ed$ and $cc$) has been already presented in Ref.\cite{co92}, 
as well as the chaining technique; so it will not be repeated here. 

In the following we shall indicate the coordinate ${\bf r}_i$ with
$i$, and the convolution integral over the coordinate $3$ as 
$ \left( | \right) $. 

The sums of the nodal diagrams $N_{xy=dd,de,ed,ee}^{\alpha \beta}$ 
are given by the solutions of the integral equations:
\[
N_{xy}^{\alpha \beta}(1,2)  =  \sum_{\gamma=p,n}\sum_{x',y'=d,e}  
\left(X_{xx'}^{\alpha \gamma}(1,3) \xi_{x'y'}^{\gamma}(3)\mid 
N_{y'y}^{\gamma \beta}(3,2) + X_{y'y}^{\gamma \beta}(3,2) 
\right) , 
\]
where $\xi^{\gamma}_{dd}=\xi^{\gamma}_d$, otherwise $\xi^{\gamma}_{xy}=\xi^{\gamma}_e$. The sum over 
$(x'y')$ runs over the pairs $(dd,de,ed)$, because of the FHNC 
convolution rules.
 
The $cc$--chains now can contain the parallel exchange link 
$\rho^\alpha_{0,P}(1,2)$ and 
the antiparallel one $\rho^\alpha_{0,A}(1,2)$. We give here  
their explicit expressions, as they represent the main changes, 
due to the j--j coupling, respect to the usual symmetric matter FHNC Eqs.
\begin{eqnarray}
N_{cc,P}^{(x)\alpha} (1,2) & = &   \left(
X_{cc,P}^{\alpha}(1,3) \xi_e^{\alpha}(3) \mid 
N_{cc,P}^{\alpha}(3,2) + X_{cc,P}^{\alpha}(3,2) - 
\rho_{0,P}^{\alpha}(3,2) 
\right) -  \nonumber \\
&  &   \left(
X_{cc,A}^{\alpha}(1,3) \xi_e^{\alpha}(3) \mid 
N_{cc,A}^{\alpha}(3,2) + X_{cc,A}^{\alpha}(3,2) - 
\rho_{0,A}^{\alpha}(3,2) 
\right)  \nonumber \\
N_{cc,P}^{(\rho)\alpha} (1,2) & = &  - \left(
\rho_{0,P}^{\alpha}(1,3) \xi_e^{\alpha}(3) \mid 
N_{cc,P}^{(x)\alpha}(3,2) + X_{cc,P}^{\alpha}(3,2) \right)  \nonumber \\
&  &   - \left(
\rho_{0,P}^{\alpha}(1,3) (\xi_e^{\alpha}(3)-1) \mid 
N_{cc,P}^{(\rho)\alpha}(3,2) - \rho_{0,P}^{\alpha}(3,2) \right)  
\nonumber \\
&  &  + \left(
\rho_{0,A}^{\alpha}(1,3) \xi_e^{\alpha}(3) \mid 
N_{cc,A}^{(x)\alpha}(3,2) + X_{cc,A}^{\alpha}(3,2) \right)   \nonumber \\
&  &   + \left(
\rho_{0,A}^{\alpha}(1,3) (\xi_e^{\alpha}(3)-1) \mid 
N_{cc,A}^{(\rho)\alpha}(3,2) - \rho_{0,A}^{\alpha}(3,2) \right) \nonumber  \\
N_{cc,A}^{(x)\alpha} (1,2) & = &   \left(
X_{cc,A}^{\alpha}(1,3) \xi_e^{\alpha}(3) \mid 
N_{cc,P}^{\alpha}(3,2) + X_{cc,P}^{\alpha}(3,2) - 
\rho_{0,P}^{\alpha}(3,2) 
\right) +  \nonumber \\
&  &   \left(
X_{cc,P}^{\alpha}(1,3) \xi_e^{\alpha}(3) \mid 
N_{cc,A}^{\alpha}(3,2) + X_{cc,A}^{\alpha}(3,2) - 
\rho_{0,A}^{\alpha}(3,2) 
\right)  \nonumber \\
N_{cc,A}^{(\rho)\alpha} (1,2) & = &  - \left(
\rho_{0,A}^{\alpha}(1,3) \xi_e^{\alpha}(3) \mid 
N_{cc,P}^{(x)\alpha}(3,2) + X_{cc,P}^{\alpha}(3,2) \right)   \nonumber \\
&  &   - \left(
\rho_{0,A}^{\alpha}(1,3) (\xi_e^{\alpha}(3)-1) \mid 
N_{cc,P}^{(\rho)\alpha}(3,2) - \rho_{0,P}^{\alpha}(3,2) \right) \nonumber  \\
&  &  - \left(
\rho_{0,P}^{\alpha}(1,3) \xi_e^{\alpha}(3) \mid 
N_{cc,A}^{(x)\alpha}(3,2) + X_{cc,A}^{\alpha}(3,2) \right)   \nonumber \\
&  &   - \left(
\rho_{0,P}^{\alpha}(1,3) (\xi_e^{\alpha}(3)-1) \mid 
N_{cc,A}^{(\rho)\alpha}(3,2) - \rho_{0,A}^{\alpha}(3,2) \right) , \nonumber
\end{eqnarray}
and 
\[
N_{cc,X}^{\alpha} (1,2) = N_{cc,X}^{(x)\alpha} (1,2) +
N_{cc,X}^{(\rho)\alpha} (1,2)  .
\]

The partial pair FHNC distribution functions, $g_{xy}$, and the 
sums of the composite diagrams, $X_{xy}$, are:
\setbox4=\hbox{$ g_{dd}^{\alpha \beta}(1,2)\biggl[ $}
\begin{eqnarray}
g_{dd}^{\alpha \beta}(1,2) & = & f^2_{\alpha \beta}(1,2) \exp
\biggl[ N_{dd}^{\alpha \beta}(1,2) + E_{dd}^{\alpha \beta}(1,2) 
\biggr] \nonumber \\
 & = & 1 + N_{dd}^{\alpha \beta}(1,2) +
X_{dd}^{\alpha \beta}(1,2)  \nonumber \\  \nonumber \\
\nonumber
g_{de}^{\alpha \beta}(1,2) & = & g_{ed}^{\beta \alpha}(2,1) \\  
& = & g_{dd}^{\alpha \beta}(1,2)
\biggl[ N_{de}^{\alpha \beta}(1,2) + E_{de}^{\alpha \beta}(1,2)
\biggr]  \nonumber \\
& = & N_{de}^{\alpha \beta}(1,2) + X_{de}^{\alpha \beta}(1,2) 
 \nonumber \\ \nonumber \\
g_{ee}^{\alpha \beta}(1,2) & = & g_{dd}^{\alpha \beta}(1,2)
\biggl[ N_{ee}^{\alpha \beta}(1,2) + E_{ee}^{\alpha \beta}(1,2)+
  \nonumber \\
& & \hskip\wd4 \left(N_{ed}^{\alpha \beta}(1,2) + E_{ed}^{\alpha \beta}(1,2)\right)
\left(N_{de}^{\alpha \beta}(1,2) + E_{de}^{\alpha
\beta}(1,2)\right)-  \nonumber \\
& &  \hskip\wd4 2 \delta_{\alpha \beta} \left(N_{cc,P}^{\alpha}(1,2) 
+ E_{cc,P}^{\alpha}(1,2) - \rho_{0,P}^{\alpha}(1,2) \right)^2-  \nonumber \\
& &   \hskip\wd4 2 \delta_{\alpha \beta} \left(N_{cc,A}^{\alpha}(1,2) 
+ E_{cc,A}^{\alpha}(1,2) - \rho_{0,A}^{\alpha} (1,2) \right)^2
\biggr]  \nonumber \\
& = & N_{ee}^{\alpha \beta}(1,2) + X_{ee}^{\alpha \beta}(1,2)  
\nonumber \\ \nonumber \\
g_{cc,X}^{\alpha}(1,2) & = & g_{dd}^{\alpha \alpha}(1,2)
\biggl[ N_{cc,X}^{\alpha}(1,2) 
+ E_{cc,X}^{\alpha}(1,2) - \rho_{0,X}^{\alpha} (1,2) \biggr]  \nonumber \\
& = & N_{cc,X}^{\alpha}(1,2) 
+ X_{cc,X}^{\alpha}(1,2) - \rho_{0,X}^{\alpha} (1,2) . \nonumber  
\end{eqnarray}
 The vertex corrections, $\xi_{x=d,e}^\alpha$, are given by:

\begin{eqnarray}
\nonumber
\xi_e^{\alpha} (1) & = & \exp \left[ U_d^{\alpha} (1) \right] \\
\xi_d^{\alpha} (1) & = &  \xi_e^{\alpha} (1) \left[ U_e^{\alpha} (1) +
\rho_0^{\alpha} (1) \right] , \nonumber 
\end{eqnarray}
 where
\setbox1=\hbox{${\displaystyle \sum_{\beta=p,n} \biggl\{ } $}
\setbox2=\hbox{${\displaystyle \sum_{\beta=p,n} \biggl\{ C_d^{\beta} 
(2) \biggl[ }$}
\setbox3=\hbox{$2 \biggl[ $}
\begin{eqnarray}
U_d^{\alpha} (1) & {\displaystyle = \int d^3 r_2 } & 
\sum_{\beta=p,n}  \biggl\{ \xi_d^{\beta} (2) \biggl[
 X_{dd}^{\alpha \beta}(1,2) - E_{dd}^{\alpha \beta}(1,2) -
 S_{dd}^{\alpha \beta}(1,2) \left( g_{dd}^{\alpha \beta}(1,2)-1 \right) 
\biggr]  +  \nonumber \\
 & &  \hskip\wd1 \xi_e^{\beta} (2) \biggl[
 X_{de}^{\alpha \beta}(1,2) - E_{de}^{\alpha \beta}(1,2) -
 S_{de}^{\alpha \beta}(1,2) \left( g_{dd}^{\alpha \beta}(1,2)-1 \right)-
 \nonumber \\
 & &  \hskip\wd2
 S_{dd}^{\alpha \beta}(1,2) g_{de}^{\alpha \beta}(1,2) 
\biggr] \biggr\}  + E_d^{\alpha} (1) \nonumber \\
U_e^{\alpha} (1) & = {\displaystyle \int d^3 r_2 \Biggl\{ } & 
\sum_{\beta=p,n} \biggl\{   
\xi_d^{\beta} (2) \biggl[
 X_{ed}^{\alpha \beta}(1,2) - E_{ed}^{\alpha \beta}(1,2) -
 S_{ed}^{\alpha \beta}(1,2) \left( g_{dd}^{\alpha \beta}(1,2)-1 \right) 
+  \nonumber \\
 & &   \hskip\wd2
 S_{dd}^{\alpha \beta}(1,2) g_{ed}^{\alpha \beta}(1,2) 
\biggr] +  \nonumber \\
 & & \hskip\wd1  \xi_e^{\beta} (2) \biggl[
 X_{ee}^{\alpha \beta}(1,2) - E_{ee}^{\alpha \beta}(1,2) -
 S_{ee}^{\alpha \beta}(1,2) \left( g_{dd}^{\alpha \beta}(1,2)-1 \right)-
 \nonumber \\
 & &  \hskip\wd2 
 S_{ed}^{\alpha \beta}(1,2) g_{de}^{\alpha \beta}(1,2) - 
 S_{de}^{\alpha \beta}(1,2) g_{ed}^{\alpha \beta}(1,2) -
 \nonumber  \\
 & &  \hskip\wd2 
 S_{dd}^{\alpha \beta}(1,2) g_{ee}^{\alpha \beta}(1,2) 
\biggr] \biggr\} + \nonumber \\
 & & 4 \xi_e^{\alpha} (2) \biggl[ S_{cc,P}^{\alpha}(1,2)
 g_{cc,P}^{\alpha}(1,2) + 
  S_{cc,A}^{\alpha}(1,2) g_{cc,A}^{\alpha}(1,2) \biggr] - \nonumber \\
 & & 2 \biggl[ \rho_{0,P}^{\alpha} (1,2) \biggl( 
 N_{cc,P}^{(\rho)\alpha}(1,2)-
 \rho_{0,P}^{\alpha}(1,2) \biggr)+\nonumber \\ 
 & & \hskip\wd3 
\rho_{0,A}^{\alpha} (1,2) \biggl( N_{cc,A}^{(\rho)\alpha}(1,2)-
 \rho_{0,A}^{\alpha}(1,2) \biggr) \biggr] \Biggr\} + 
E_e^{\alpha} (1) , \nonumber
\end{eqnarray}
with: 
\[
S_{xy}^{\alpha \beta} (1,2) = E_{xy}^{\alpha \beta} (1,2)+ (1/2)
N_{xy}^{\alpha \beta} (1,2)
\]
Notice that $\rho_1^\alpha(1)=\xi_{d}^\alpha(1)$.

\section*{Appendix B}

Here we will shortly present the FHNC equations for the one--body 
density matrices $\rho^{ss'\alpha}(1,1')$. These are modifications of 
the equations presented  in Ref.\cite{co94}. As we have discussed 
throught the text we have to separate the case with the same
third component of spin 
\[
\rho^{ss\alpha}(1,1') = - \xi_\omega^\alpha (1) \xi_\omega^\alpha (1')
g_{\omega\omega}^{\alpha}(1,1') 
\bigl[ N^{\alpha}_{\omega_c\omega_c,P}(1,1') +
 E^{\alpha}_{\omega_c\omega_c,P}(1,1') - 
 \rho_{0,P}^{\alpha}(1,1') \bigr] ,
\]
from the case with different ones
\[
\rho^{s-s\alpha}(1,1') = - \xi_\omega^\alpha (1) \xi_\omega^\alpha (1')
g_{\omega\omega}^{\alpha}(1,1') 
\bigl[ N^{\alpha}_{\omega_c\omega_c,A}(1,1') +
E^{\alpha}_{\omega_c\omega_c,A}(1,1') - \rho_{0,A}^{\alpha}(1,1') \bigr] .
\]
$g_{\omega\omega}^{\alpha}$ is defined as:
\[
g_{\omega\omega}^\alpha(1,1')  = 
\exp\biggl[ N_{\omega\omega}^\alpha(1,1') + E_{\omega\omega}^\alpha(1,1')
\biggr] .
\]
The nodal functions 
$N^{\alpha}_{\omega\omega}, N^{\alpha}_{\omega_c\omega_c,X}$ are 
solutions of convolution equations, formally identical to those 
presented in appendix A, with the substitution of the external $(1,2)$ 
link indices $(d,c)$ by $(\omega,\omega_c)$, where needed. New nodal
terms $N^{\alpha\beta}_{\omega d,\omega e},
N^{\alpha\beta}_{\omega_c c,X}$ 
appear in these convolutions. The same rule as before applies to them.

New partial FHNC distribution functions need to be introduced. They 
are given by:
\setbox4=\hbox{$ g_{dd}^{\alpha \beta}(1,2)\biggl[ $}
\begin{eqnarray}
g_{\omega d}^{\alpha \beta}(1,2) & = &g_{d \omega}^{\beta
\alpha}(2,1) \nonumber \\
& = & f_{\alpha \beta}(1,2) \exp
\biggl[ N_{\omega d}^{\alpha \beta}(1,2) + E_{\omega d}^{\alpha \beta}(1,2) 
\biggr] \nonumber \\
 & = & 1 + N_{\omega d}^{\alpha \beta}(1,2) +
X_{\omega  d}^{\alpha \beta}(1,2)  \nonumber \\  \nonumber \\
g_{\omega  e}^{\alpha \beta}(1,2) & = & 
g_{e \omega }^{\beta \alpha}(2,1)  \nonumber \\  
& = & g_{\omega  d}^{\alpha \beta}(1,2)
\biggl[ N_{\omega  e}^{\alpha \beta}(1,2) 
+ E_{\omega  e}^{\alpha \beta}(1,2)
\biggr]  \nonumber \\
& = & N_{\omega  e}^{\alpha \beta}(1,2) + 
X_{\omega  e}^{\alpha \beta}(1,2) 
 \nonumber \\ \nonumber \\
g_{\omega_c c,X}^{\alpha \alpha}(1,2) & =  & 
g_{\omega d}^{\alpha \alpha}(1,2)
\biggl[ N_{\omega_cc,X}^{\alpha\alpha}(1,2) 
+ E_{\omega_cc,X}^{\alpha\alpha}(1,2) - 
\rho_{0,X}^{\alpha} (1,2) \biggr]  \nonumber \\
& = & N_{\omega_cc,X}^{\alpha\alpha}(1,2) 
+ X_{\omega_cc,X}^{\alpha\alpha}(1,2) - 
\rho_{0,X}^{\alpha} (1,2) . \nonumber  
\end{eqnarray}

The vertex corrections, $\xi_\omega^\alpha$, is:
\[
\xi_\omega^{\alpha} (1)  =  \exp \left[ U_\omega^{\alpha} (1) \right] , 
\]
with:
\setbox1=\hbox{${\displaystyle \sum_{\beta=p,n} \biggl\{ } $}
\setbox2=\hbox{${\displaystyle \sum_{\beta=p,n} \biggl\{ C_d^{\beta} 
(2) \biggl[ }$}
\setbox3=\hbox{$2 \biggl[ $}
\begin{eqnarray}
U_\omega^{\alpha} (1) & {\displaystyle = \int d^3 r_2 } & 
\sum_{\beta=p,n}  \biggl\{ \xi_d^{\beta} (2) \biggl[
 X_{\omega d}^{\alpha \beta}(1,2) - E_{\omega d}^{\alpha \beta}(1,2) -
 S_{\omega d}^{\alpha \beta}(1,2) \left( 
 g_{\omega d}^{\alpha \beta}(1,2)-1 \right) 
\biggr]  +  \nonumber \\
 & &  \hskip\wd1 \xi_e^{\beta} (2) \biggl[
 X_{\omega e}^{\alpha \beta}(1,2) - E_{\omega e}^{\alpha \beta}(1,2) -
 S_{\omega e}^{\alpha \beta}(1,2) \left( 
 g_{\omega d}^{\alpha \beta}(1,2)-1 \right)-
 \nonumber \\
 & &  \hskip\wd2
 S_{\omega d}^{\alpha \beta}(1,2) g_{\omega e}^{\alpha \beta}(1,2) 
\biggr] \biggr\}  + E_\omega^{\alpha} (1) . \nonumber 
\end{eqnarray}

\section*{Appendix C}

In this appendix, we shall give the explicit expressions for the
uncorrelated one--body densities and related quantities. 
For the s.p. wavefunctions we use:
\[
\phi^{\alpha}_{nljm} (\vec{x}_i) \ = \ R_{nlj}^\alpha (r_i) \sum_{\mu,s}
\langle l \mu \onehalf s \mid j m \rangle Y_{l \mu} (\hat{r}_i) \chi_s (i)
\chi_\alpha (i)
\]

The one--body density  $\rho_{T1}^\alpha$ is given by:
\[
\rho_{T1}^\alpha ({\bf r}_1) = \frac 1 {4 \pi} \sum_{nlj} (2j+1) \left[
R_{nlj}^\alpha (r_1) \left( D_{nlj}^\alpha(r_1) - \frac{l(l+1)}{r_1^2}
R_{nlj}^\alpha (r_1) \right) -(R_{nlj}^{\alpha} (r_1)')^2 \right]
\]
where we have defined:
\[
D_{nlj}^\alpha(r_1)= \nabla_1^2 R_{nlj}^{\alpha} (r_1) 
=R_{nlj}^{\alpha\prime\prime} (r_1) +
\frac 2 {r_1} R_{nlj}^{\alpha\prime} (r_1) - \frac{l(l+1)}{r_1^2}
R_{nlj}^{\alpha} (r_1) .
\]

For the one--body density matrix $\rho_{T2}^\alpha$  we find:
\begin{eqnarray}
\rho_{T2}^\alpha({\bf r}_1,{\bf r}_2)& = & \frac 1 {2(4\pi)^2}
\sum_{nlj \atop n'l'j'} (2j+1)(2j'+1)R_{nlj}^\alpha (r_2)
R_{n'l'j'}^\alpha (r_2) \nonumber \\
& & \Biggl\{ \left[ R_{nlj}^\alpha (r_1)D_{n'l'j'}^\alpha (r_1)-
R_{nlj}^{\alpha\prime} (r_1)R_{n'l'j'}^{\alpha\prime} (r_1) \right]
P_l(\cos \theta)P_{l'}(\cos \theta)-  \nonumber \\
& & \frac{\sin^2 \theta}{r_1^2}
R_{nlj}^{\alpha} (r_1)R_{n'l'j'}^{\alpha} (r_1)
P_l^\prime(\cos \theta)P_{l'}^\prime(\cos \theta) \Biggr\}+
 \nonumber \\
& & \frac 2 {(4\pi)^2}
\sum_{nlj \atop n'l'j'} (-1)^{j+j'-l-l'-1}
R_{nlj}^\alpha (r_2) R_{n'l'j'}^\alpha (r_2) \nonumber \\
& & \Biggl\{ \left[ R_{nlj}^\alpha (r_1)D_{n'l'j'}^\alpha (r_1)-
R_{nlj}^{\alpha\prime} (r_1)R_{n'l'j'}^{\alpha\prime} (r_1) \right]
Q_l(\cos \theta)Q_{l'}(\cos \theta)- \nonumber \\
& & \frac{\sin^2(\theta)}{r_1^2}
R_{nlj}^{\alpha} (r_1)R_{n'l'j'}^{\alpha} (r_1)
Q_l^\prime(\cos \theta) Q_{l'}^\prime(\cos \theta) \Biggr\}  , \nonumber  
\end{eqnarray}
where $P_l$ are Legendre polynomials, $\theta$ is the angle between
the vectors ${\bf r}_1$ and ${\bf r}_2$ and  we have
defined:
\begin{eqnarray}
Q_l (\cos \theta ) & = & \sin \theta P_l^\prime (\cos \theta) \nonumber \\
Q_l^\prime (\cos \theta ) & = & \frac 1 {\sin \theta}
\left( \cos \theta  P_l^\prime (\cos \theta)
- l(l+1) P_l (\cos \theta) \right) . \nonumber 
\end{eqnarray}

For the $T_3$ densities we obtain:
\begin{eqnarray}
\rho_{T3,P}^\alpha({\bf r}_1,{\bf r}_2)& = & 2 \nabla_1^2
\rho_{0,P}^\alpha({\bf r}_1,{\bf r}_2)= \frac 1 {4\pi}
\sum_{nlj} (2j+1)R_{nlj}^\alpha (r_2)D_{nlj}^\alpha (r_1)
P_l(\cos \theta) \nonumber \\
\rho_{T3,A}^\alpha({\bf r}_1,{\bf r}_2)& = & 2 \nabla_1^2
\rho_{0,A}^\alpha({\bf r}_1,{\bf r}_2)= \frac 1 {2\pi}
\sum_{nlj} (-1)^{j-l-1/2}R_{nlj}^\alpha (r_2)D_{nlj}^\alpha (r_1)
Q_l (\cos \theta) . \nonumber 
\end{eqnarray}

Finally, $\rho_{T4}^\alpha$, appearing in the calculation of the 
center of mass term $T_{cm}$, is given by:
\[
\rho_{T4}^\alpha({\bf r}_1,{\bf r}_2) =
\rho_{T6}^\alpha({\bf r}_1,{\bf r}_2) -
\rho_{0,P}^\alpha({\bf r}_1,{\bf r}_2) 
\rho_{T5,P}^\alpha({\bf r}_1,{\bf r}_2)-
\rho_{0,A}^\alpha({\bf r}_1,{\bf r}_2)
\rho_{T5,A}^\alpha({\bf r}_1,{\bf r}_2) , 
\]
where 
\begin{eqnarray}
\rho_{T6}^\alpha({\bf r}_1,{\bf r}_2) & = & 2 \left(
\nabla_1 \rho_{0,P}^\alpha({\bf r}_1,{\bf r}_2) \cdot
\nabla_2 \rho_{0,P}^\alpha({\bf r}_1,{\bf r}_2) +
\nabla_1 \rho_{0,A}^\alpha({\bf r}_1,{\bf r}_2) \cdot
\nabla_2 \rho_{0,A}^\alpha({\bf r}_1,{\bf r}_2) \right) , \nonumber \\
\rho_{T5,X}^\alpha({\bf r}_1,{\bf r}_2) & = & 2 
\nabla_1 \cdot \nabla_2 \rho_{0,X}^\alpha({\bf r}_1,{\bf r}_2) , 
\nonumber 
\end{eqnarray}
and
\begin{eqnarray}
\rho_{T5,P}^\alpha({\bf r}_1,{\bf r}_2)& = & \frac 1 {4\pi}
\sum_{nlj} (2j+1) \Biggl[ R_{nlj}^{\alpha\prime}(r_1)
R_{nlj}^{\alpha\prime} (r_2)
\cos \theta P_{l}(\cos \theta)+ \nonumber \\
& & \left( 
R_{nlj}^{\alpha} (r_2) \frac{R_{nlj}^{\alpha} (r_1)}{r_1}+
R_{nlj}^{\alpha} (r_1) \frac{R_{nlj}^{\alpha} (r_2)}{r_2} \right)
\sin^2 \theta P_l^\prime(\cos \theta)+
 \nonumber \\
& &  \frac{R_{nlj}^{\alpha} (r_1)R_{nlj}^{\alpha} (r_2)}{r_1r_2}
\left(\sin^2 \theta P_l^\prime(\cos \theta)+
l(l+1) \cos \theta P_{l}(\cos \theta)\right) \Biggr]  \nonumber \\
\nonumber 
\rho_{T5,A}^\alpha({\bf r}_1,{\bf r}_2)& = & \frac 1 {2\pi}
\sum_{nlj} (-1)^{j-l-1/2} \Biggl[ R_{nlj}^{\alpha\prime}(r_1)
R_{nlj}^{\alpha\prime} (r_2) \cos \theta Q_{l}(\cos \theta)+ \nonumber \\
& & \left( 
R_{nlj}^{\alpha} (r_2) \frac{R_{nlj}^{\alpha} (r_1)}{r_1}+
R_{nlj}^{\alpha} (r_1) \frac{R_{nlj}^{\alpha} (r_2)}{r_2} \right)
\sin^2 \theta Q_l^\prime(\cos \theta) \nonumber \\
& & \frac{R_{nlj}^{\alpha} (r_1)R_{nlj}^{\alpha} (r_2)}{r_1r_2} 
\left\{ \sin^2 \theta Q_{l}^\prime(\cos \theta)+ \right. \nonumber \\
& & \left. \left( l(l+1)-
\frac 1 {\sin^2 \theta} \right) \cos \theta Q_l(\cos \theta) \right\}
 \Biggr] \nonumber \\ 
\rho_{T6}^\alpha({\bf r}_1,{\bf r}_2)& = & \frac 1 {2(4\pi)^2}
\sum_{nlj \atop n'l'j'} (2j+1)(2j'+1)R_{nlj}^\alpha (r_2)
R_{n'l'j'}^\alpha (r_1) \nonumber \\
& & \Biggl\{ \cos \theta \Biggl[ R_{nlj}^{\alpha\prime} (r_2)
R_{n'l'j'}^{\alpha\prime} (r_1)
P_l(\cos \theta)P_{l'}(\cos \theta)-  \nonumber \\
& & \sin^2 \theta \frac{R_{nlj}^{\alpha} (r_2)R_{n'l'j'}^{\alpha} (r_1)}
{r_1 r_2} P_l^\prime(\cos \theta)P_{l'}^\prime(\cos \theta)
\Biggr]+ \nonumber \\
& &  \sin^2 \theta \Biggl[
\frac{R_{nlj}^{\alpha} (r_2)}{r_2}R_{n'l'j'}^{\alpha\prime} (r_1)
 P_l^\prime(\cos \theta)P_{l'}(\cos \theta)+ \nonumber \\
 & & R_{nlj}^{\alpha\prime} (r_2)\frac{R_{n'l'j'}^{\alpha} (r_1)}
{r_1} P_l(\cos \theta)P_{l'}^\prime(\cos \theta) \Biggr] \Biggr\}
 \nonumber \\
& & \frac 2 {(4\pi)^2}
\sum_{nlj \atop n'l'j'} (-1)^{j+j'-l-l'-1}R_{nlj}^\alpha (r_2)
R_{n'l'j'}^\alpha (r_1) \nonumber \\
& & \Biggl\{ \cos \theta \Biggl[ R_{nlj}^{\alpha\prime} (r_2)
R_{n'l'j'}^{\alpha\prime} (r_1)
f_l(\cos \theta)f_{l'}(\cos \theta)-  \nonumber \\
& & \sin^2 \theta \frac{R_{nlj}^{\alpha} (r_2)R_{n'l'j'}^{\alpha} (r_1)}
{r_1 r_2} f_l^\prime(\cos \theta)f_{l'}^\prime(\cos \theta)
\Biggr]+ \nonumber \\
& &  \sin^2 \theta \Biggl[
\frac{R_{nlj}^{\alpha} (r_2)}{r_2}R_{n'l'j'}^{\alpha\prime} (r_1)
 f_l^\prime(\cos \theta)f_{l'}(\cos \theta)+ \nonumber \\
 & & R_{nlj}^{\alpha\prime} (r_2)\frac{R_{n'l'j'}^{\alpha} (r_1)}
{r_1} f_l(\cos \theta)f_{l'}^\prime(\cos \theta) \Biggr] \Biggr\} . 
\nonumber 
\end{eqnarray}

\newpage
\section*{Tables}

\vspace*{1.0cm}
\begin{center}
\begin{tabular}{|l|l|rrr|}
\hline
& q & $a^{\alpha \beta}_q$ & $b^{\alpha \beta}_q$ & $c^{\alpha \beta}_q$ \\
\hline
                   & 1 &  1 &  1 &  1 \\
                   & 2 &  0 &  3 & -1 \\
$\alpha  =  \beta$ & 3 &  1 &  1 &  1 \\
                   & 4 &  0 &  3 & -1 \\
\hline
                   & 1 &  1 &  0 &  0 \\
                   & 2 &  0 &  0 &  0 \\
$\alpha \ne \beta$ & 3 & -1 &  2 &  2 \\
                   & 4 &  0 &  6 & -2 \\
\hline
\end{tabular}
\end{center}
{\bf Table 1.} Coefficients of eq.(\ref{rho2q}).

\vspace*{2.0cm}
\begin{center}
\begin{tabular}{|l|rrrrr|}
\hline
   & $^{12}$C & $^{16}O$ & $^{40}$Ca & $^{48}$Ca & $^{208}$Pb \\
\hline
  $S_p$          & 0.999 & 1.000 & 1.000 & 1.000 & 0.995 \\
  $S_n$          & 0.999 & 1.000 & 1.000 & 0.999 & 0.996 \\
  $S_{pp}$       & 0.994 & 0.998 & 1.000 & 0.999 & 0.989 \\
  $S_{np}$       & 0.998 & 1.000 & 1.000 & 0.999 & 0.991 \\
  $S_{nn}$       & 0.994 & 0.997 & 1.000 & 0.998 & 0.993 \\
  $S_{\sigma,0}$ & 1.004 & 1.018 & 1.009 & 1.014 & 0.953 \\
  $S_{\sigma,1}$ & 1.001 & 1.000 & 1.000 & 1.002 & 0.985 \\
\hline
\end{tabular}
\end{center}
{\bf Table 2.} Ground state sum rules. $S_{\sigma,0(1)}$ are the ratios 
$S_\sigma^{corr}/S_\sigma^{unc}$ in FHNC/0 and FHNC--1 approximations, 
respectively (see text).

\vskip 2.cm
\begin{center}
\begin{tabular}{|c|ccc|}
\hline
     &  $V_0$  & $R$ & $a$ \\
\hline
$^{12}C$   & 52.5 & 3.20  & 0.53   \\
$^{16}O$   & 52.5 & 3.20  & 0.53   \\
$^{40}Ca$  & 57.5 & 4.10  & 0.53   \\
$^{48}Ca$  & 57.5 & 4.10  & 0.53   \\
$^{208}Pb$ & 60.4 & 7.46  & 0.79   \\
\hline
\end{tabular}
\end{center}
{\bf Table 3.} Coefficients of the Woods--Saxon potential (\ref{wood}) used
for the calculations of Tab.4. 
%
\newpage
\vskip 1.cm
\begin{center}

\begin{tabular}{|c|rrrr|}
\hline
$^{12}C$  &  F1    & F2        &  F3  & F4   \\
\hline
     $V$  & -21.28 & -21.43  & -21.43  & -21.24   \\
   $V_C$  &   0.00 &   0.00  &   0.64  &   0.63   \\
     $T$  &  19.10 &  19.28  &  19.28  &  19.09   \\
     $E$  &  -2.18 &  -2.15  &  -1.51  &  -1.52   \\
\hline
$^{16}O$  &  F1    & F2   & F3  &  F4   \\
\hline
     $V$  & -26.18 & -26.18 & -26.18 & -25.84   \\
   $V_C$  &   0.00 &   0.00 &   0.87 &   0.86   \\
     $T$  &  20.16 &  20.16 &  20.16 &  19.87   \\
     $E$  &  -6.02 &  -6.02 &  -5.15 &  -5.11   \\
\hline
$^{40}Ca$  &  F1    & F2   & F3  &  F4   \\
\hline
     $V$  & -32.61  & -32.61 & -32.61  & -31.97   \\
   $V_C$  &   0.00  &   0.00 &   1.95  &   1.91   \\
     $T$  &  23.96  &  23.96 &  23.96  &  23.45   \\
     $E$  &  -8.65  &  -8.65 &  -6.70  &  -6.61   \\
\hline
$^{48}Ca$  &  F1    & F2   & F3  &  F4   \\
\hline
     $V$  & -33.41  & -33.69 & -33.69  & -32.13   \\
   $V_C$  &   0.00  &   0.00 &   1.62  &   1.59   \\
     $T$  &  25.77  &  26.12 &  26.12  &  25.68   \\
     $E$  &  -7.64  &  -7.57 &  -5.95  &  -5.86   \\
\hline
$^{208}Pb$  &  F1    & F2   & F3  &  F4   \\
\hline
     $V$  & -33.44  & -33.70 & -33.70  & -32.67  \\
   $V_C$  &   0.00  &   0.00 &   3.98  &   3.84  \\
     $T$  &  24.04  &  24.29 &  24.29  &  23.68  \\
     $E$  &  -9.40  &  -9.41 &  -5.43  &  -5.15  \\
\hline
\end{tabular}

\end{center}

{\bf Table 4.} Energies per nucleon, in $MeV$, for the five nuclei 
considered. These results have been obtained using the Woods--Saxon
parameters of Tab.3, the S3 interaction of Afnan and Tang \cite{afn68} and
the Euler ACA correlation. The healing distances 
are $d=2.1 fm$ for $^{12}$C and $d=2.0 fm$ for the other nuclei. 

\vskip 2.cm
\begin{center}
\begin{tabular}{|c|c|cccc|}
\hline
         &  &  $V_0$ & $V_{LS}$  & $R$ & $a$ \\
\hline
$^{12}$C   & p  & 62.00 & 3.20 & 2.86 & 0.57   \\
           & n  & 60.00 & 3.15 & 2.86 & 0.57   \\
\hline
$^{16}$O   & p  & 52.50 & 7.00 & 3.20 & 0.53   \\
           & n  & 52.50 & 6.54 & 3.20 & 0.53   \\
\hline
$^{40}$Ca  & p  & 57.50 & 11.11 & 4.10 & 0.53   \\
           & n  & 55.00 & 8.50 & 4.10 & 0.53   \\
\hline
$^{48}$Ca  & p  & 59.50 & 8.55 & 4.36 & 0.53   \\
           & n  & 50.00 & 7.74 & 4.36 & 0.53   \\
\hline
$^{208}$Pb & p  & 60.40 & 6.75 & 7.46 & 0.79   \\
           & n  & 44.32 & 6.08 & 7.46 & 0.66   \\
\hline
\end{tabular}
\end{center}
{\bf Table 5.} Coefficients of the Woods--Saxon potential (\ref{wood}). 

\vskip 2. cm
\begin{center}
\begin{tabular}{|l|l|cccccccc|}
\hline
 &      & $<V>$ & $<V_c>$ & $T_\phi$ & $T_F$ & $T_{cm}$ & $<H>$ & $ E/A $ & d \\
\hline
         & $EU$  & -335.5 & 8.2 & 228.7 & 65.4 & 12.8 & -33.2 & -3.84 & 2.44 \\
$^{12}$C & $ACA$ & -291.9 & 8.2 & 223.2 & 45.0 & 12.8 & -15.5 & -2.36 & 2.20 \\
         & $G$   & -273.2 & 8.0 & 221.3 & 31.3 & 12.8 & -12.6 & -2.12 &      \\
\hline
         & $EU$  & -489.5 & 14.4 & 273.3 & 82.1 & 11.4 & -119.7 & -8.20 & 2.36 \\
$^{16}$O & $ACA$ & -418.5 & 13.9 & 266.8 & 55.0 & 11.4 & -82.8 & -5.89 & 2.08 \\
         & $G$   & -395.8 & 13.5 & 263.2 & 43.4 & 11.4 & -75.7 & -5.44 &   \\
\hline
         & $EU$  & -1515.1 & 80.4 & 791.3 & 262.6 & 10.1 &-381.2 & -9.78 & 2.42
\\  
$^{40}$Ca & $ACA$ & -1302.2 & 78.0 & 771.9 & 190.2 & 10.1 & -262.1 & -6.81 &
2.20 \\
         & $G$    & -1220.0 & 75.6 & 750.3 & 154.2 & 10.1 & -239.9 & -6.25 & \\
\hline
         & $EU$  & -1744.9 & 78.8 & 968.8 & 302.4 & 9.6 & -394.9 & -8.43 & 2.45 \\
$^{48}$Ca & $ACA$ & -1494.5 & 76.0 & 936.6 & 209.6 & 9.6 & -272.3 & -5.87 & 2.15
\\
         & $G$   & -1418.8 & 74.3 & 918.7 & 182.7 & 9.6 & -243.2 & -5.27 & \\
\hline
     & $EU$  & -7600.3 & 810.8 & 3865.3 & 1163.6 & 6.0 & -1761.4 & -8.50 & 2.52
\\ $^{208}$Pb & $ACA$ & -6772.0 & 809.2 & 3910.7 & 995.8 & 6.0 & -1056.3 & -5.11 &
2.45 \\
         & $G$   & -6218.4 & 792.1 & 3740.7 & 814.1 & 6.0 & -871.5 & -4.22 & \\
\hline
\end{tabular}
\end{center}
{\bf Table 6.} Ground state expectation values with different correlations. 
$EU$ indicates the Euler correlation, ACA the ACA Euler correlation and
$G$ the ACA gaussian correlation.  
The parameters of the gaussian correlations are $A=0.7$
for all the nuclei considered and $B=2.15 fm^{-2}$ for $^{48}$Ca,
$B=2.2 fm^{-2}$ for $^{12}$C, $^{40}$Ca and $^{208}$Pb, and $B=2.3 fm^{-2}$
for  $^{16}$O. $T_\phi=T^{(1)}_\phi+T^{(2)}_\phi+T^{(3)}_\phi$.
Energies in $MeV$ and distances in $fm$.

\vskip 2.cm
\begin{center}
\begin{tabular}{|l|cccccccc|}
\hline
 &      $<V>$ & $<V_c>$ & $T_\phi$ & $T_F$ & $T_{cm}$ & $<H>$ & $ E/A $ & d \\
\hline
$^{12}$C & -293.6 & 8.1 & 221.9 & 15.6 & 12.8 & -48.0 & -5.07 & 2.17 \\
\hline
$^{16}$O  & -435.0 & 13.6 & 252.4 & 20.9 & 11.4 & -136.7 & -9.26 & 2.07 \\
\hline
$^{40}$Ca  & -1328.6 & 76.7 & 759.2 & 78.2 & 10.1 & -414.5  & -10.62 & 2.25 \\
\hline
$^{48}$Ca  & -1554.1 & 75.6 & 931.1 & 93.2 & 9.6 & -454.2 & -9.66 & 2.30
\\ 
\hline
$^{208}$Pb  & -7017.8 & 805.3 & 3827.5 & 390.2 & 6.0 & -1994.8 & -9.62 &
2.50 \\ 
\hline
\end{tabular}
\end{center}
\vspace*{1.0cm}
{\bf Table 7.} Ground state energies for the ACA Euler correlation and the
B1 potential. Energies in $MeV$ and distances in $fm$.

\vskip 2. cm
\begin{center}
\begin{tabular}{|l|cc|cc|}
\hline
 &        $S3$  &         & $B1$       &       \\
\hline
 &       $E_1$ & $\Delta$ & $E_1$    & $\Delta$  \\
\hline
$^{12}$C   & -2.21 & 6.36 & -4.63 &  8.68 \\
$^{16}$O   & -5.57 & 5.43 & -8.61 &  7.02  \\
$^{40}$Ca  & -6.33 & 7.04 & -10.05 &  5.37  \\
$^{48}$Ca  & -5.28 & 10.05 & -9.33 & 3.42  \\
$^{208}$Pb & -5.14 & -0.59 & -9.69 & -0.73 \\
\hline
\end{tabular}
\end{center}
\vspace*{1.0cm}
{\bf Table 8.} FHNC--1 binding energies per nucleon calculated with the
Euler--ACA correlations for the S3 and B1 interactions. 
$\Delta$ is the percentile deviation from the FHNC/0  energy.

\newpage
\vskip 2. cm
\begin{center}
\begin{tabular}{|l|l|cc c|}
\hline
     &   & $MD^p_0/x_p$ &  $MD^n_0/x_n$ & $MD_2/<T>$ \\ 
\hline
          & $EU$  & 0.990 & 0.990 & 0.955  \\
$^{12}$C  & $ACA$ & 0.995 & 0.995 & 0.960  \\
          & $G$   & 0.997 & 0.997 & 0.942   \\
\hline
          & $EU$  & 0.996 & 0.996 & 0.969  \\
$^{16}$O  & $ACA$ & 0.996 & 0.996 & 0.963  \\
          & $G$   & 0.996 & 0.996 & 0.954   \\
\hline
          & $EU$  & 1.000 & 0.995 & 0.956 \\  
$^{40}$Ca & $ACA$ & 0.995 & 0.996 & 0.942 \\
          & $G$   & 0.992 & 0.993 & 0.919  \\
\hline
          & $EU$  & 0.998 & 0.993 & 0.949 \\
$^{48}$Ca & $ACA$ & 0.993 & 0.994 & 0.936 \\
          & $G$   & 0.991 & 0.993 & 0.914  \\
\hline
           & $EU$  & 0.999 & 1.005 & 0.943  \\
$^{208}$Pb & $ACA$ & 0.996 & 1.002 & 0.934 \\
           & $G$   & 0.984 & 0.985 & 0.917  \\
\hline
\end{tabular}
\end{center}
{\bf Table 9.} Momentum distributions SRs, Eqs.(\ref{sumnk},\ref{sumknk}), for the
calculation of Tab.6. $x_p=Z/A$ and $x_n=N/A$ are the proton and neutron 
concentrations respectively.

\newpage
{\bf Figure captions}

\vskip 1. cm
\noindent
{\bf Fig.1.} ACA Euler (upper panel) and gaussian (lower panel) 
correlation functions at the energy minimum for several nuclei, 
for the S3 potential.

\vskip 1. cm
\noindent
{\bf Fig.2.} Euler and ACA Euler correlation functions at the energy 
minimum for $^{12}$C and $^{48}$Ca nuclei with the S3 interaction.
Full lines are the ACA Euler correlations. The lowest panel gives 
the nuclear matter $nn(=pp)$ (dot--dashed) and $np$ (dashed) Euler 
correlations fot the same potential, at $d=2.0 fm$.

\vskip 1. cm
\noindent
{\bf Fig.3.} Proton density distributions. Full lines are the
uncorrelated densities. The dotted curves are obtained with the ACA
gaussian correlation, the dashed curves with the ACA Euler correlation and
the dot--dashed curves with the isospin dependent Euler ones. The S3
interaction has been used. 
The dash-doubly-dotted curves are the densities obtained 
with the B1 potential using the ACA Euler correlation.

\vskip 1. cm
\noindent
{\bf Fig.4.} Proton momentum distributions. 
Curves as in Fig.3. 

\vskip 1. cm
\noindent
{\bf Fig. 5.} Proton density (upper panel)  and momentum distribution
(lower panel) for $^{208}$Pb. 
Curves as in Fig.3.

\vskip 1. cm
\noindent
{\bf Fig. 6.} ACA Euler correlation functions for the B1
interaction.

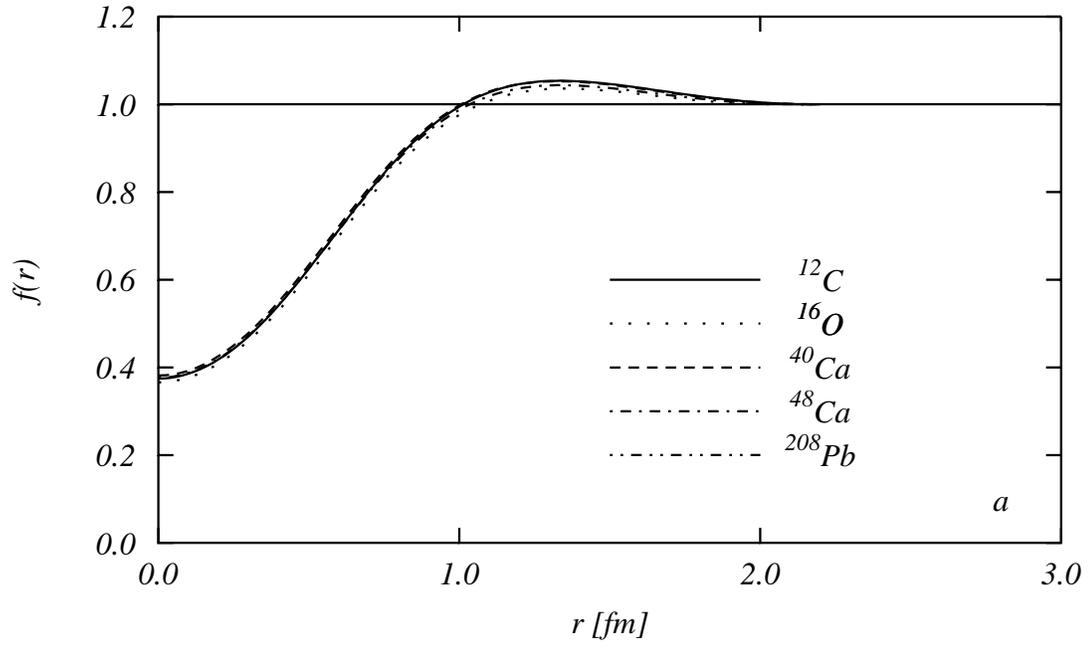
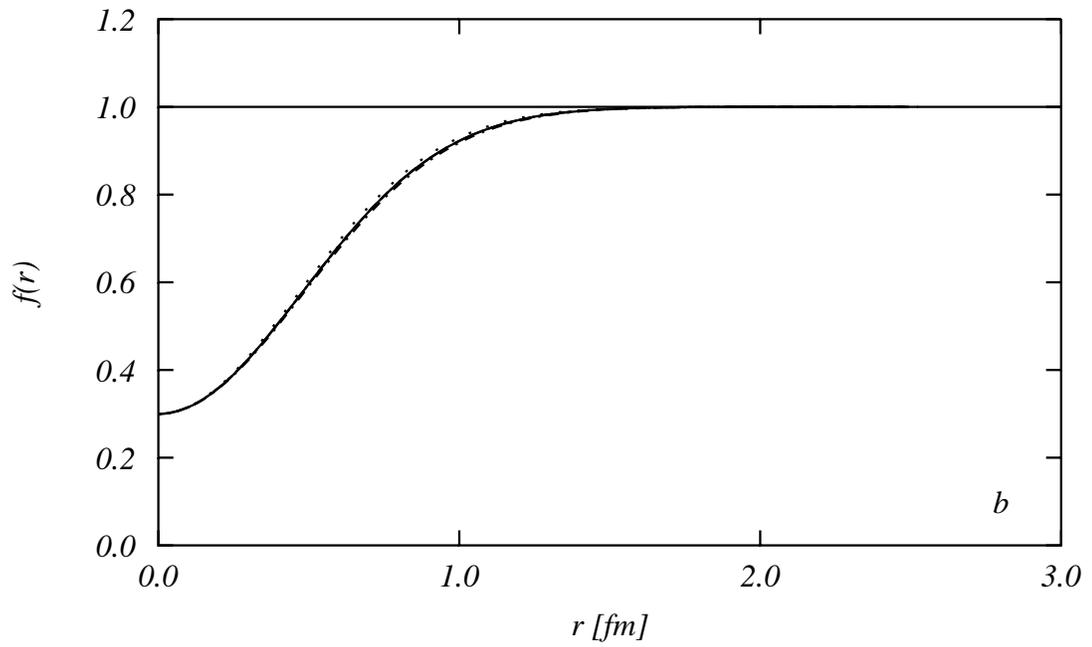
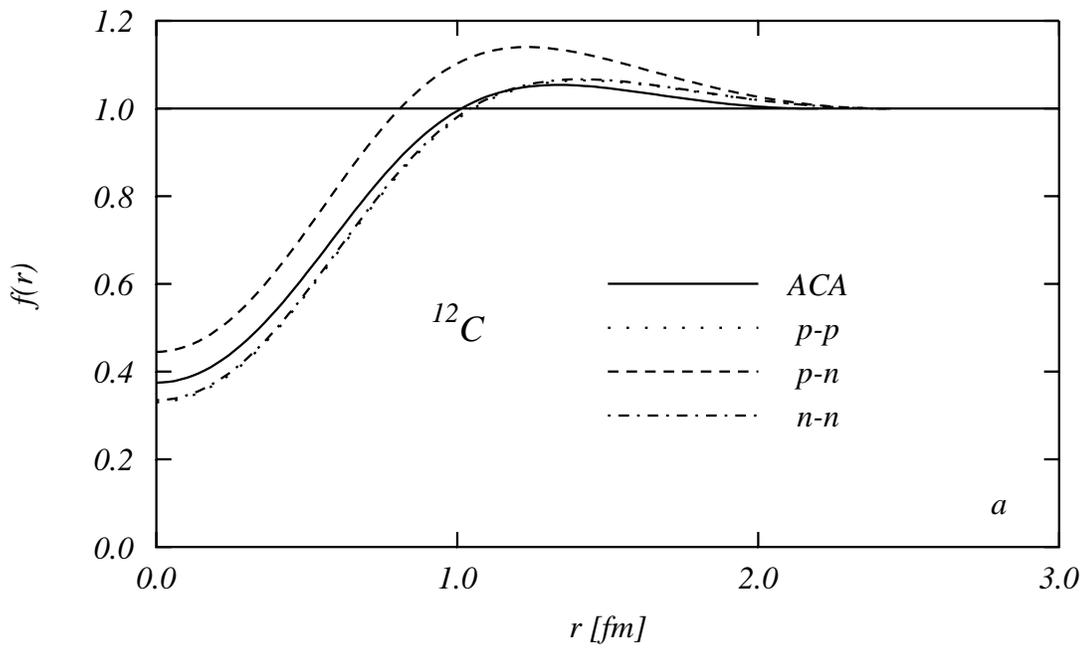
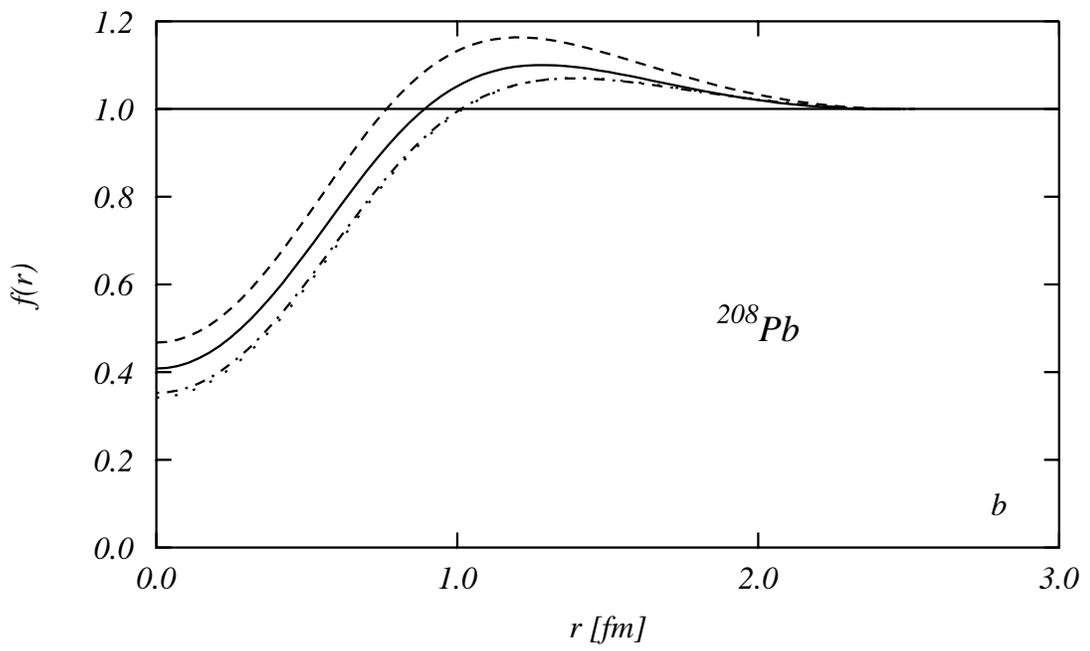
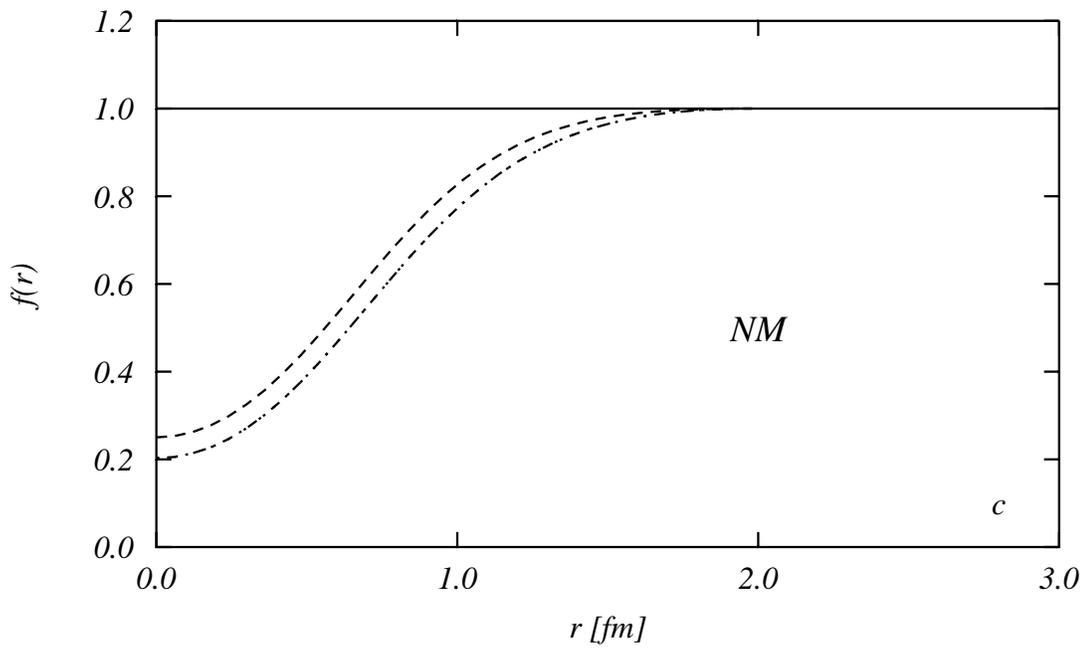
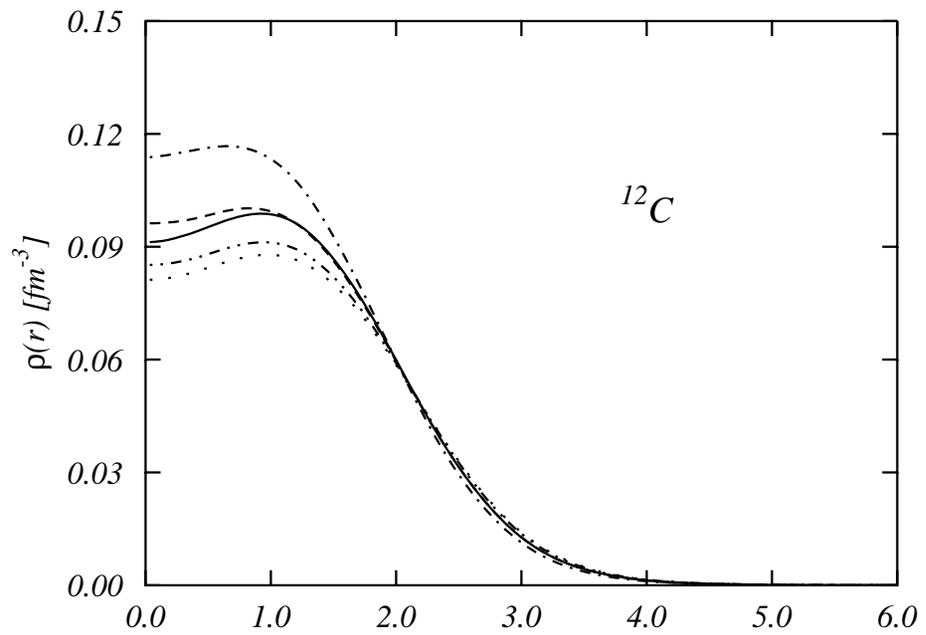
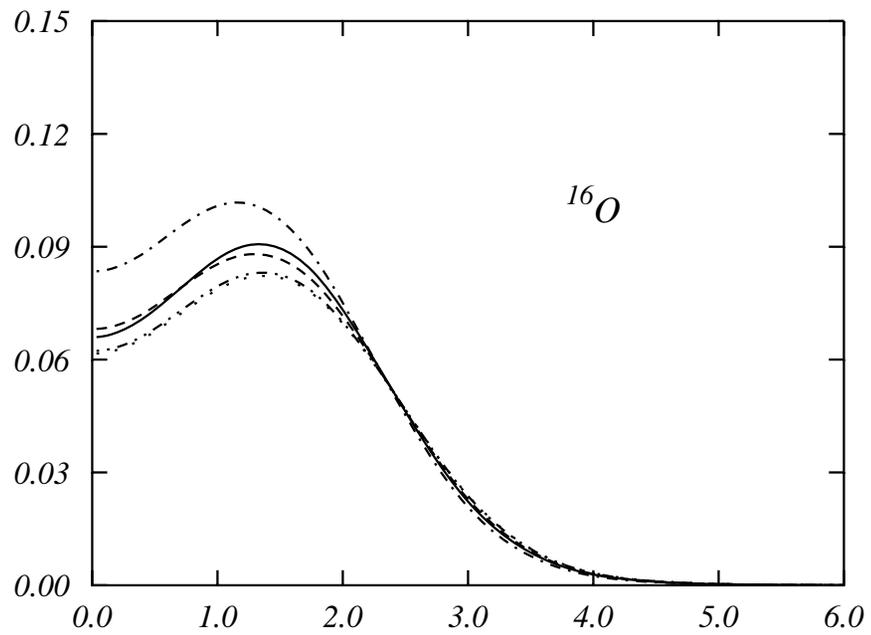
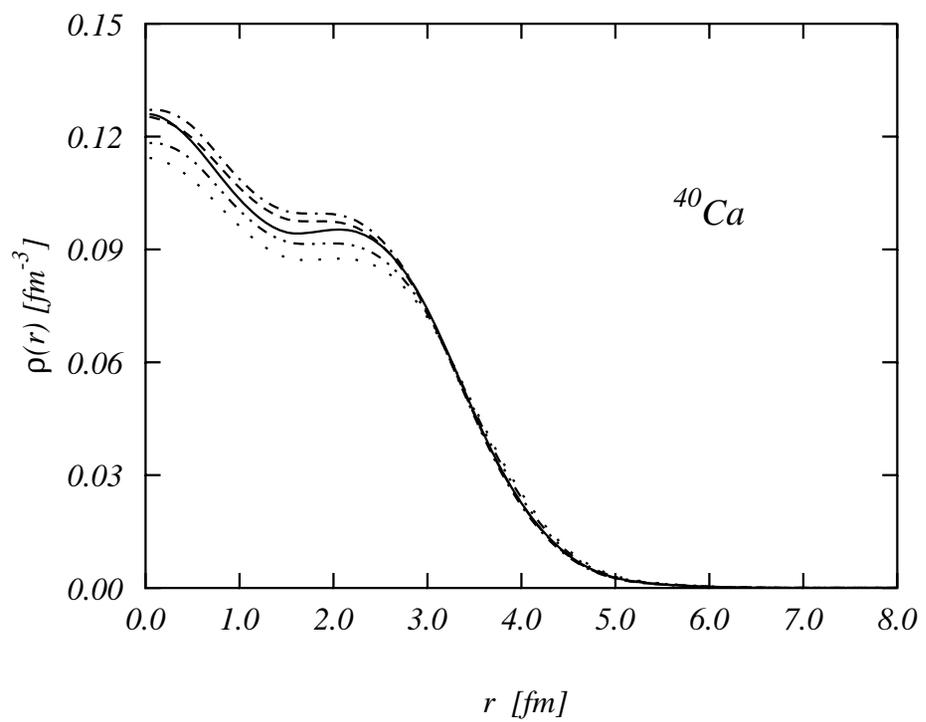
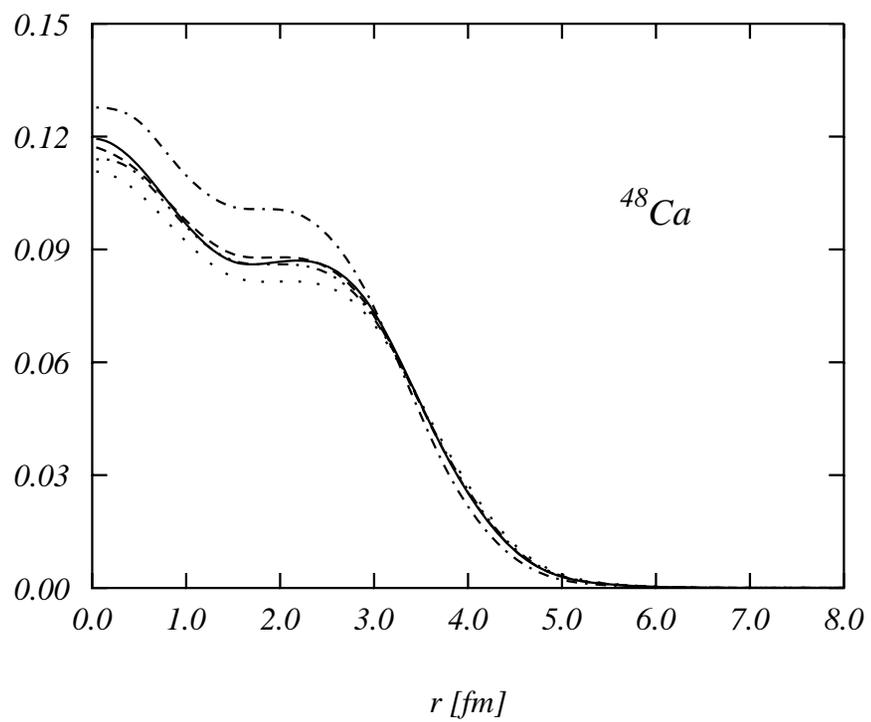
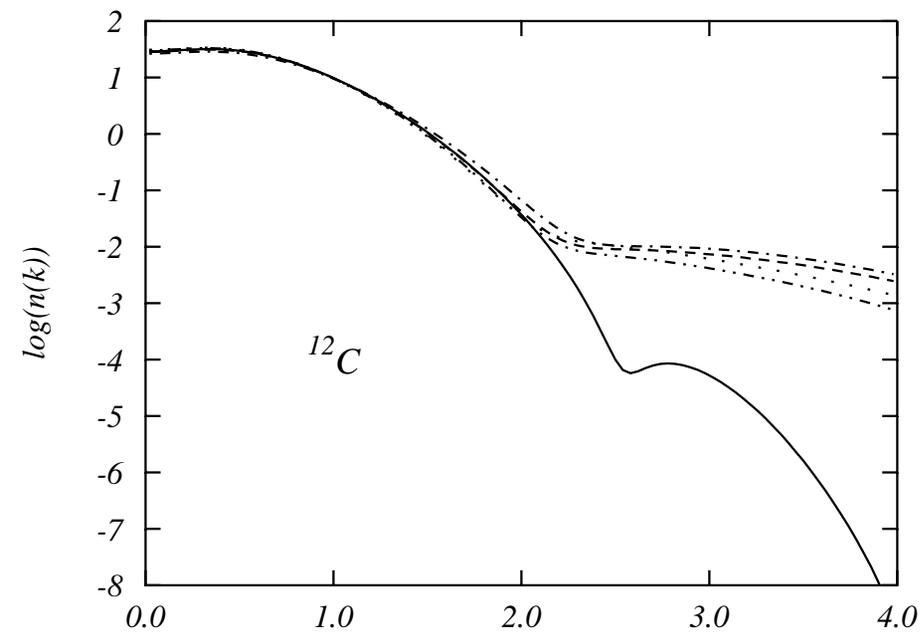
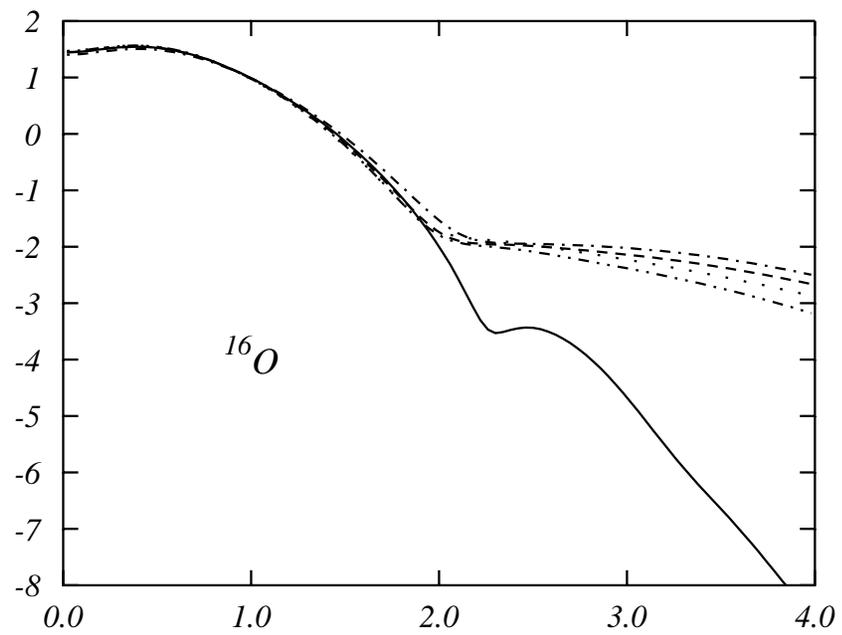
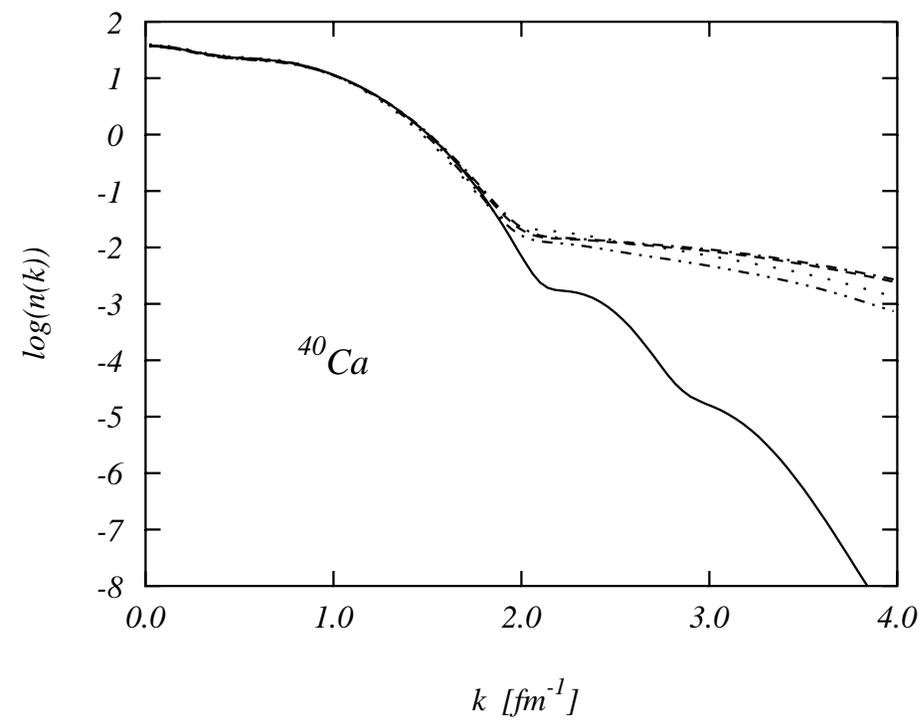
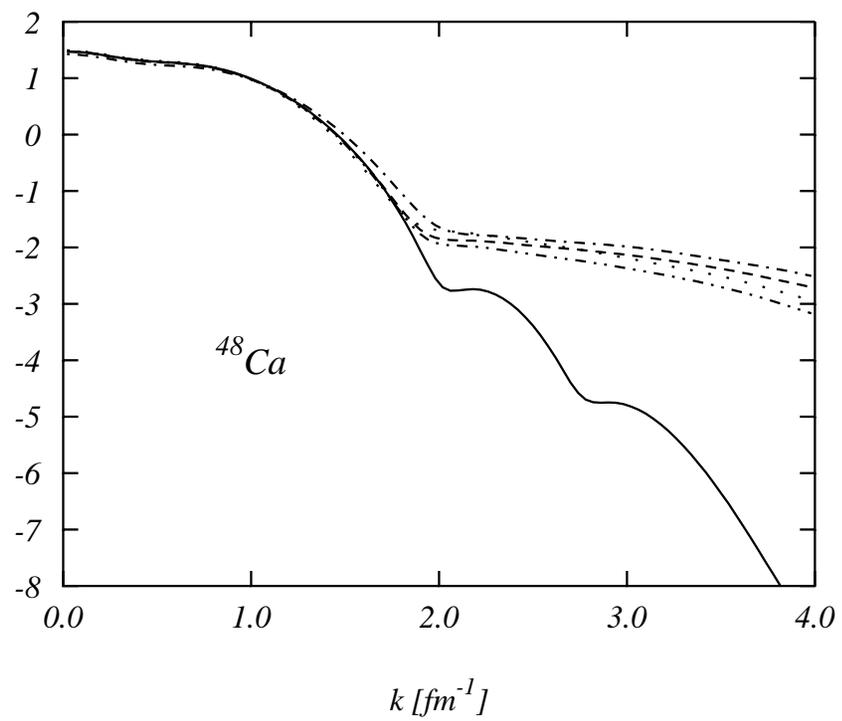
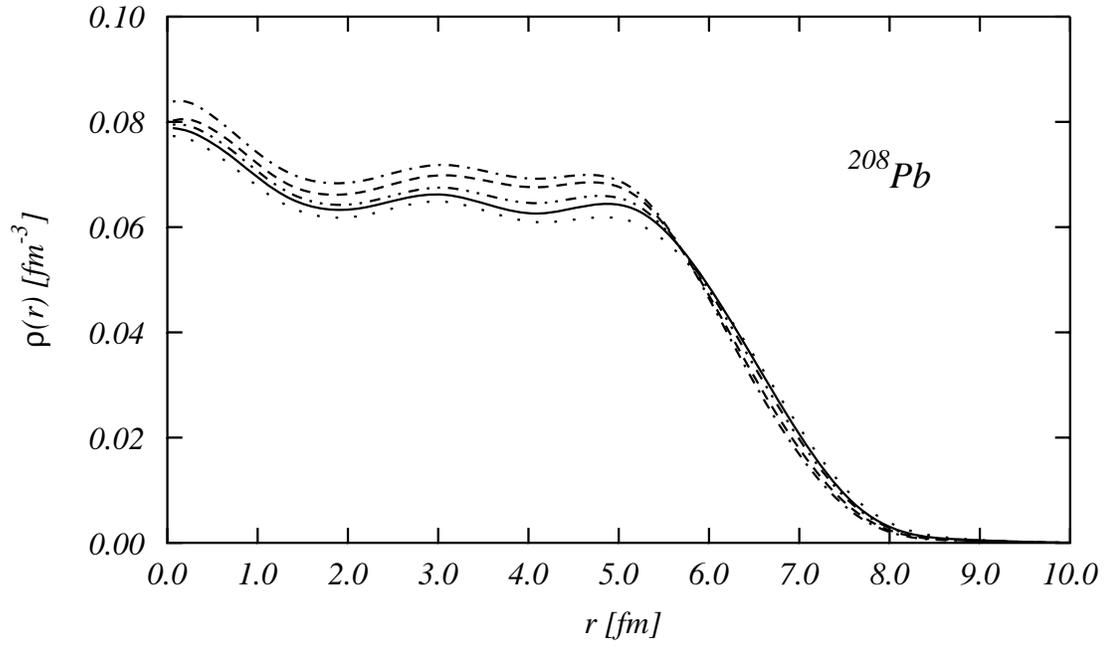
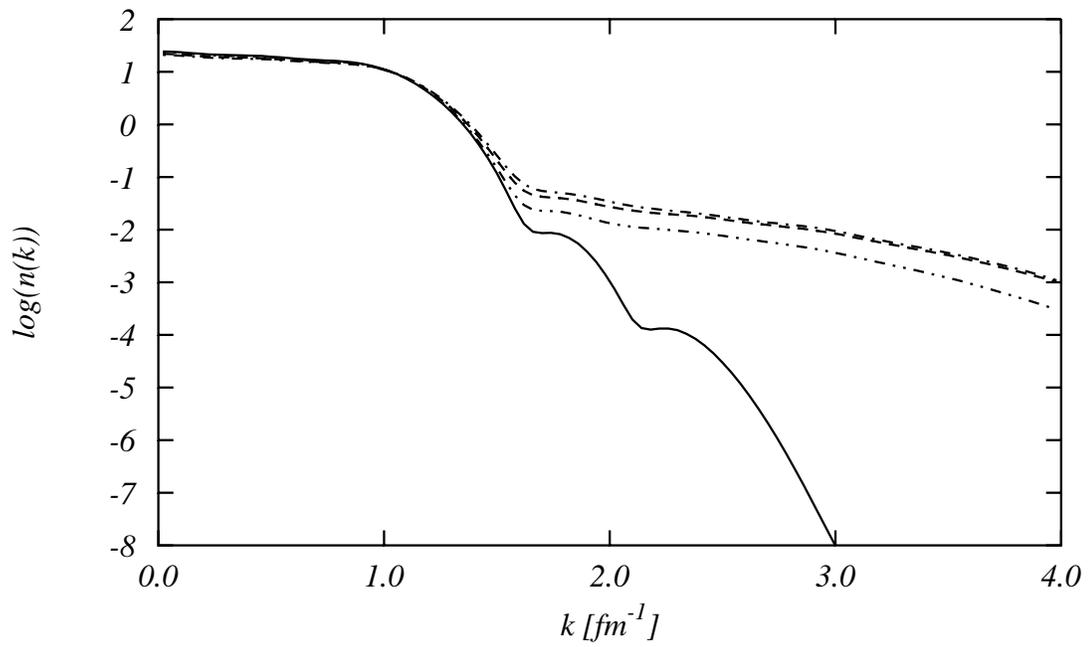
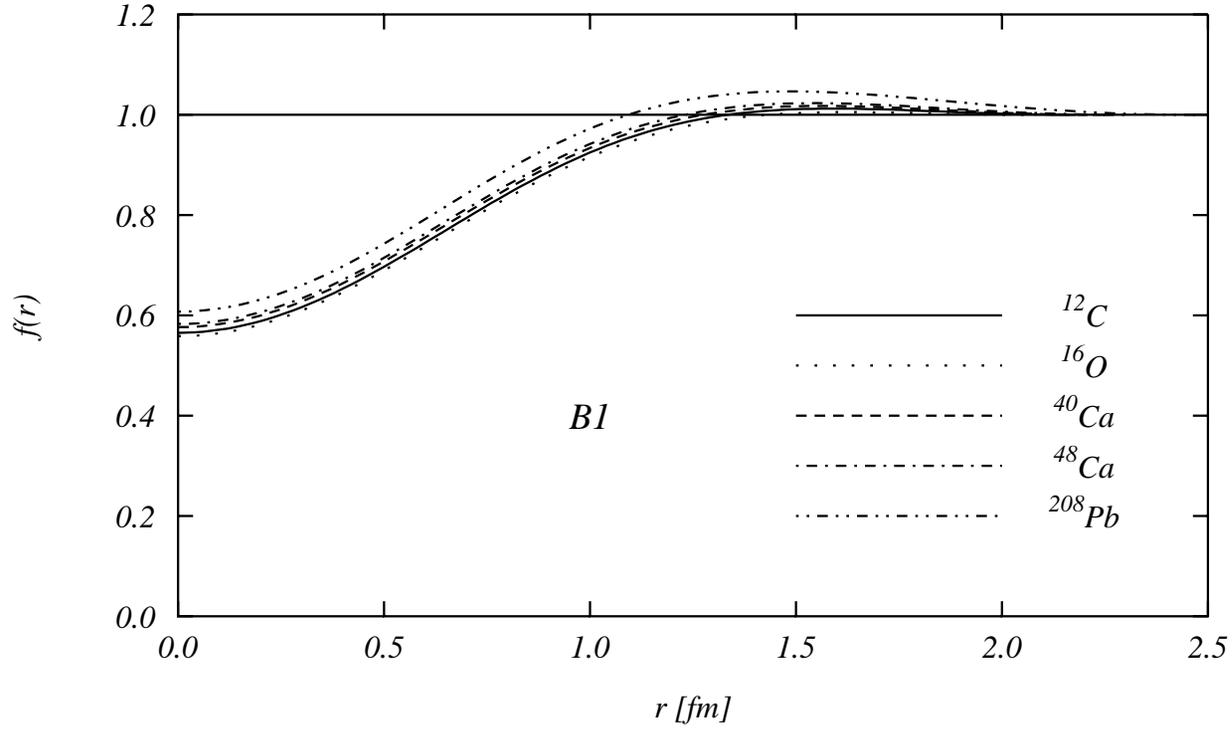
\end{document}